\documentclass{aastex63}

\usepackage{natbib,lineno}
\linenumbers

\shorttitle{Self-Consistent Atmosphere Representation and Interaction}
\shortauthors{Peterson et al.}

\begin{document}
\nolinenumbers

\title{Self-Consistent Atmosphere Representation and Interaction in Photon Monte Carlo Simulations}

\author{J.~R.~Peterson, G.~Sembroski, A.~Dutta, C.~Remacaldo}

\affil{Department of Physics and Astronomy, Purdue University, West Lafayette, IN 47907, USA}

\correspondingauthor{John~R. ~Peterson}
\email{peters11@purdue.edu}

\begin{abstract}
  We present a self-consistent representation of the atmosphere and implement the interactions of light with the atmosphere using a photon Monte Carlo approach.  We compile global climate distributions based on historical data, self-consistent vertical profiles of thermodynamic quantities, spatial models of cloud variation and cover, and global distributions of four kinds of aerosols.  We then implement refraction, Rayleigh scattering, molecular interactions, Tyndall-Mie scattering to all photons emitted from astronomical sources and various background components using physics first principles.   This results in emergent image properties that include: differential astrometry and elliptical point spread functions predicted completely to the horizon, arcminute-scale spatial-dependent photometry variations at 20 mmag for short exposures, excess background spatial variations at 0.2\% due the atmosphere, and a point spread function wing due to water droplets.   We use a common atmophere representation framework to self-consistently model all phenomena by simulating individual photons.
  We reproduce the well-known correlations in image characteristics:  correlations in altitude with absolute photometry (overall transmission) and relative photometry (spectrally-dependent transmission), anti-correlations of altitude with differential astrometry (non-ideal astrometric patterns) and background levels, and an anti-correlation in absolute photometry with cloud depth.  However, we also find further subtle correlations including an anti-correlation of temperature with background and differential astrometry, a correlation of temperature with absolute and relative photometry, an anti-correlation of absolute photometry with humidity, a correlation of humidity with Lunar background, a significant correlation of PSF wing with cloud depth, an anti-correlation of background with cloud depth, and a correlation of lunar background with cloud depth.
\end{abstract}


\section{Introduction}

Simulations have become increasingly indispensible in modern astrophysics.  In particular, a wide range of simulation approaches have been pursued in X-ray astronomy (\citealt{peterson2004}; \citealt{peterson2007}; \citealt{andersson2007}; \citealt{davis2012}) and optical astronomy (\citealt{ackermann2012}; \citealt{lane}, \citealt{ellerbroek}; \citealt{lelouarn}; \citealt{britton}; \citealt{jolissaint}; \citealt{bertin}; \citealt{dobke}; \citealt{peterson2015}, \citealt{rowe2015}).  In our previous work, we have created an ab initio physics simulator (PhoSim) that follows individual photons through the atmosphere telescope and camera (\citealt{peterson2015}, \citealt{peterson2019}, \citealt{burke2019a}, \citealt{peterson2019}).  The detailed microphysics of photon and electron interactions are followed to predict the final pixel (or removal) of individual photons.  Recent work has extended the first principles physics to include the deformation physics of optics (\citealt{peterson2019}) and the electrostatic physics of sensors (\citealt{peterson2020}).

A number of publications have demonstrated a diverse set of applications for an ab initio simulator in astronomy using PhoSim (https://www.phosim.org).  PhoSim has been used to:  1) to test the design of Rubin/LSST (\citealt{angeli2016}, \citealt{thomas2016} \citealt{xin2015}) and JWST (\citealt{burke2019a}) 2) to plan for future observations (e.g. \citealt{merlin2023}, \citealt{bretonniere2023}, \citealt{sanchez2020}, \citealt{thomas2018} \citealt{bard2016}, \citealt{mandelbaum2014}, \citealt{bard2013} \citealt{chang2013a}, \citealt{chang2013b}, \citealt{chang2012}), 3) for advanced image processing algorithm development (e.g. \citealt{nie2021a}, \citealt{nie2021b}, \citealt{carlsten2018}, \citealt{li2016} \citealt{meyers2015}), 4) to understand physical effects (\citealt{walter2015}, \citealt{beamer2015}), and 5) for advanced AI/machine learning development by simulating training sets (\citealt{burke2019b}).  In addition to published studies, hundreds of users have used PhoSim for informal studies.  Most commonly the applications include:  1) simulations to plan for future observations, 2) testing analysis algorithms on a simulation with perfect information, and 3) developing the designs of future telescopes.  Several terabytes of images have been produced for a variety of data challenges for multiple observatories.  These applications have also led to the community collectively validate the physics implementation in PhoSim.

It is increasingly important to consider the full self-consistent representation of the objects while simulating the photon (and electron) propagation and interactions.  This is important since emergent effects that are imprinted on images, including the point-spread-function (PSF) size and shape and the photometric and astronometric patterns, are directly affected in correlated ways.  This is difficult since it involves simulating the ab initio physics within a common framework and does not allow individual effects to be treated in isolation.  Unfortunately, that leads to a much more complicated implentation.   Then, it is a difficult to build more detail without the simulator both becoming opaque to the user that contains hidden assumptions and degenerating into one of many possible models for a particular aspect of the simulation.  However, in previous work with PhoSim, we have left the modeling aspect of the simulation (i.e. the instrument design details) exposed as a tiered set of user inputs, and built the simulator entirely around the physical response of the photons and subsequent electrons.  In that way, the implementation of the latter would be unambiguous and follow ab initio physics.

We apply this concept in this work with the simulation of a self-consistent representation of the atmosphere.  Then, the photon interactions are applied unambigously on that particular representation of the atmosphere.  However, the simulator is built so that representation of the atmosphere can also be modified or simplified by the user.  The observational features are not affected by the representation of the atmosphere because they only depend on the photon interactions, but it is critical that the photon interactions are simulated in a self-consistent basis with the same atmosphere representation in order to reproduce correlations of observables.  Therefore, we first compile a global representation of the atmosphere based on a number of measurements in the literature in \S2.  The subsequent photon interactions with the self-consistent representation are described in \S3.   The resulting observational effects are described in \S4, and the correlated patterns are outlined in \S5.  Conclusions and future work is described in \S6.  We limit the scope to the non-turbulent aspect of the atmosphere as the turbulent atmosphere can be handled as in \citealt{peterson2015} with one exception.

This approach improves over standard methologies in three key ways.  The first is that we compile a complete global representation of the atmosphere.  We use theoretical and empirical descriptions of the atmosphere to describe thermodynamic quantities for every location on the globe.  As discussed above, although the reference implementation may not be perfect and it depends on the current fidelity of the atmosphere modelling in the literature, the response of light to the represented atmosphere follows exact first principles physics.  Having a complete global description is increasingly important in the PhoSim package since dozens of ground-based telescopes are implemented, and it lowers the burden to the user to add additional telescopes.  This leads to the ability to compare image of different observatories across the globe in a straight-forward manner.

The second improvement is that emergent phenomena are uniquely predicted with no additional parameters.  In particular, both the background emission pattern and the point-spread-function wing due to the atmosphere are self-consistently predicted along with layer transmission absorption functions and differential chromatic refraction.  In a photon by photon approach through segments of the atmosphere, the emergent observables are predicted.  Thus, observables such as the differential refraction pattern, the background level, point-spread-function wing, and transmission properties are uniquely predicted from the choice of atmosphere representation.

Finally, the third improvement is by using a physics-based approach some of the common approximations and simplification are not used in favor of an exact approach.  For instance, given that refraction affects the path length of light through the atmosphere and that the atmospheric composition varies with altitude, means that using a single airmass quantity is not appropriate.  Similarly, in some cases the transmission of light calculations assume that light is lost when it is actually scattered and still present in the image.  Additionally, the curvature of the Earth affects transmission and refraction patterns at large zenith angles.  In most cases, these complications are actually solved in a straight-forward manner with a photon Monte Carlo approach, despite being difficult to implement with analytic approaches.

We expect that the detailed implementation in this work may be insufficient to predict exact long-term atmosphere patterns for a given site.  That application is best studied with specific observations.  However, there are many applications to predict realistic astronomical images that could be obtained at a given site and this work combines ab initio physics and careful representations cover the globe.  In particular, three common uses are:  1) detailed simulation of a telescope while it is being designed, built, and commissioned to understand scientific performance and explain complex phenomena, 2) planning of future observations for a given telescope before the data is obtained, 3) simulation of realistic images with perfect input knowledge to test any complex image processing pipelines with associated algorithms, and 4) simulation of realistic training sets for machine learning/artificial intelligence applications.  We also estimate in the conclusion future and current validation.

\section{Input Self-Consistent Atmosphere Representation}

In this section, we develop the framework for the input to PhoSim of a self-consistent atmosphere representation.  We will need to describe the relative quantity of different constituents at different locations both as a function of altitude and across the globe.  This includes molecules, water droplets, and aerosols, including the size distributions of water droplets and aerosols.  Similarly, we will self-consistently represent thermodynamic quantities, such as temperature and pressure.  Various parts of this will be used for the photon interactions used in \S3.   The most important feature is that we use the same atmosphere representation for all of the photon interactions.  We also allow any aspect of this representation to be modified by the user.  For example, in \S4 we consider modifying the ground temperature but then letting the temperaure as a function of altitude be determined from the input data below.

The complete description of the atmosphere is a very complex problem where there is no current theoretical solution.  Therefore, we will rely on a combination of empirical historical data augmented with theoretical profiles where there is insufficient detail.  The accuracy required is not simple to define, since we will use the thermodynamic quantities and molecular and particle abundances to predict a number of interactions including transmission, refraction, and scattering in \S3.  However, we will use as comprehensive representation as possible that will reasonably approximate the range of atmospheric conditions.

\subsection{Climate Historical Data}

\begin{figure}[htb]
\begin{center}
\includegraphics[width=0.99\columnwidth]{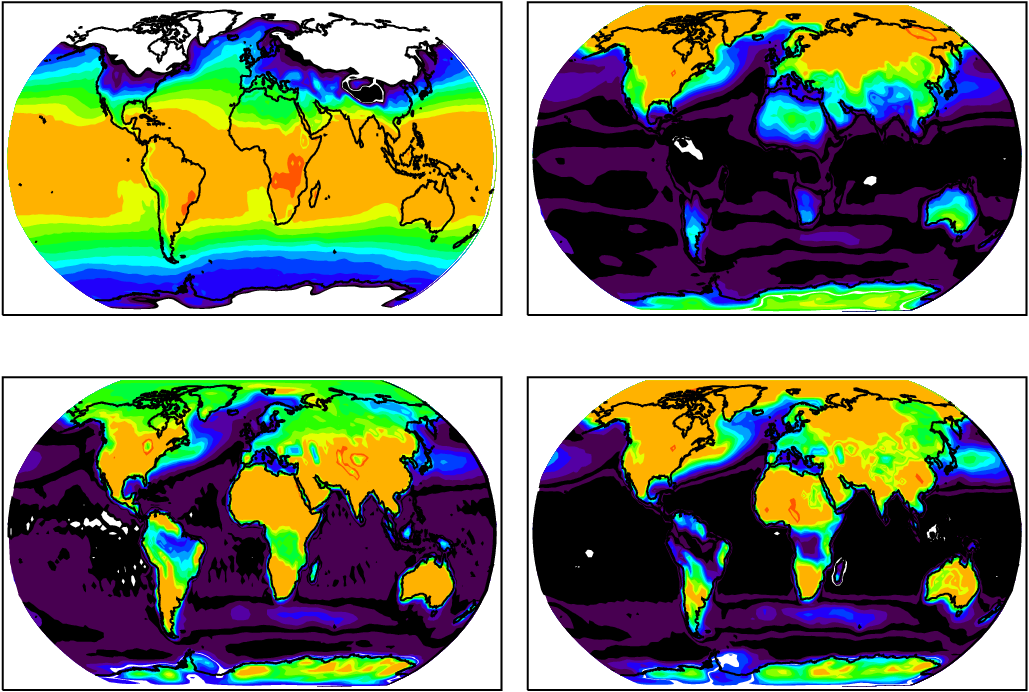}
\end{center}
\caption{\label{fig:label1} Temperature maps across the globe for the month of January.  The top left shows the mean temperature, the top right is the standard deviation for that month, the bottom left shows the mean diurnal range, and the bottom right shows the standard deviation of the diurnal range.  We compute the statistics using the data from \cite{kalnay1996} as described in the text.  The scales go from purple to red from 257 to 296 K, 0.50 to 5.7 K, 0.44 to 5.7 K, and 0.44 to 2.7 K for the plots, respectively.}
\end{figure}

\begin{figure}[htb]
\begin{center}
\includegraphics[width=0.99\columnwidth]{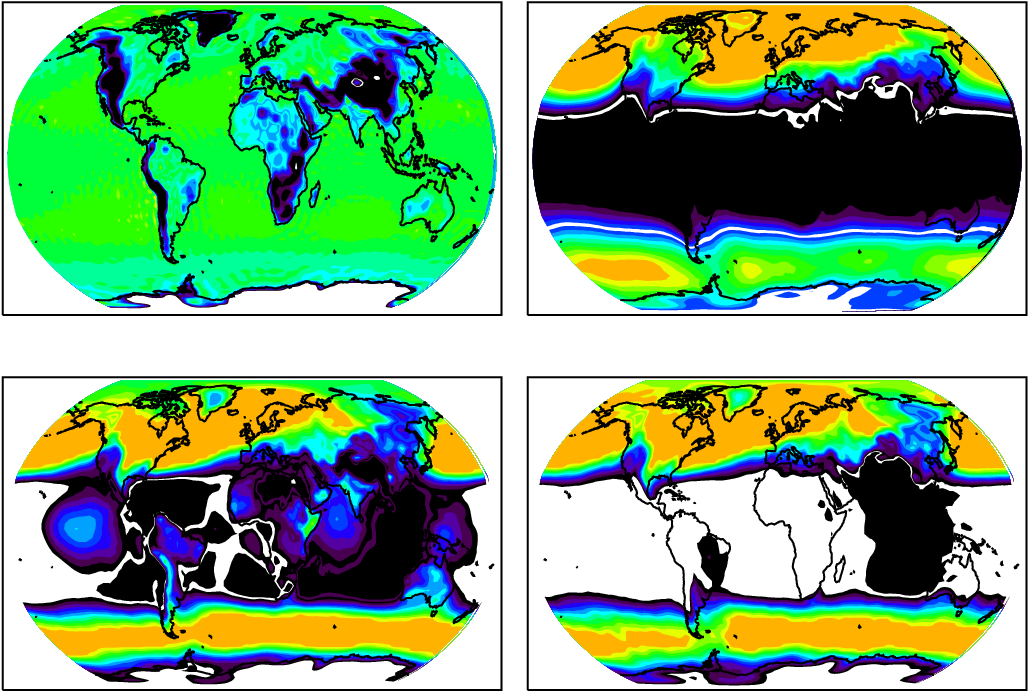}
\end{center}
\caption{\label{fig:label2} Pressure maps across the globe for the month of January.  The top left shows the mean pressure, the top right is the standard deviation for that month, the bottom left shows the mean diurnal range, and the bottom right shows the standard deviation of the diurnal range.  We compute the statistics using the data from \cite{kalnay1996} as described in the text.   The scales go from purple to red from 87000 to 105963 Pa, 270 to 1135 Pa, 269 to 622 Pa, and 130 to 440 K for the plots, respectively.}
\end{figure}

\begin{figure}[htb]
\begin{center}
\includegraphics[width=0.99\columnwidth]{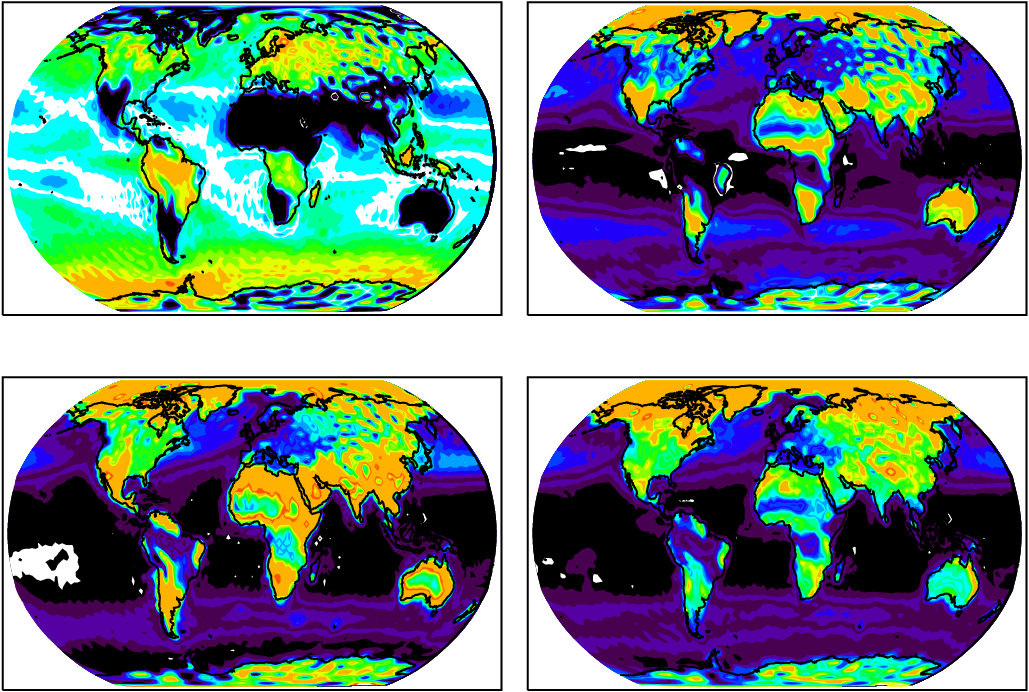}
\end{center}
\caption{\label{fig:label3} Relative humidity maps across the globe for the month of January.  The top left shows the mean humidity, the top right is the standard deviation for that month, the bottom left shows the mean diurnal range, and the bottom right shows the standard deviation of the diurnal range.  We compute the statistics using the data from \cite{kalnay1996} as described in the text.   The scales go from purple to red from 66 to 94 \%, 3.0 to 15 \%, 5.7 to 25 \%, and 2.7 to 16 \% for the plots, respectively.}
\end{figure}

\begin{figure}[htb]
\begin{center}
\includegraphics[width=0.99\columnwidth]{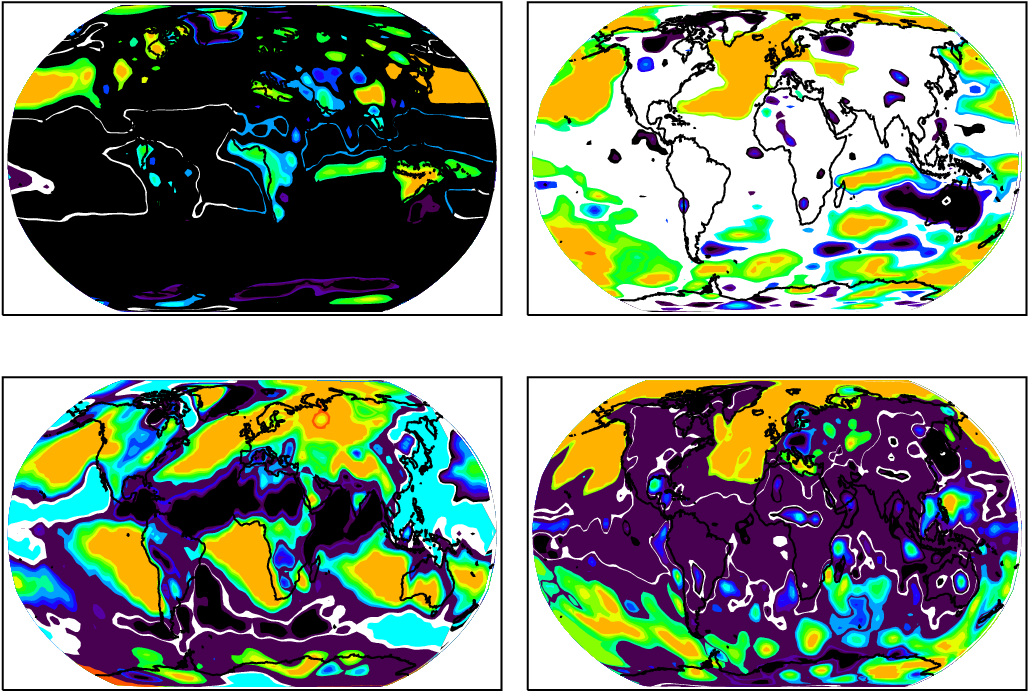}
\end{center}
\caption{\label{fig:label4} Wind patterns across the globe at high altitude consisting of a pressure of 10 mb in January.  The top left shows the mean longitudinal velocity component, the top right shows the variance of the logitudinal velocity component, the bottom left shows the mean latitudinal velocity component, and the bottom right shows the variance of the latitudinal velocity component.  We compute the statistics using the data from \cite{kalnay1996} as described in the text.   The scales go from purple to red from -3.9 to 4.0 m/s, 0.85 ot 2.5 m/s, -2.6 to 2.2 m/s, and 0.77 to 2.1 m/s for the plots, respectively.}
\end{figure}

\begin{figure}[htb]
\begin{center}
\includegraphics[width=0.99\columnwidth]{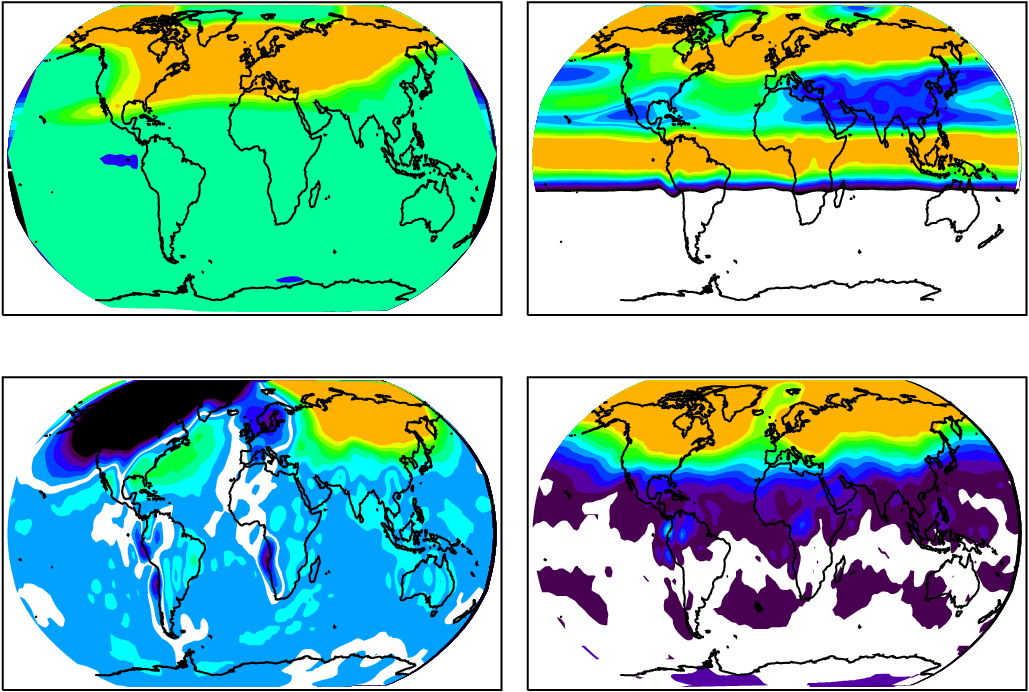}
\end{center}
\caption{\label{fig:label5} Wind patterns across the globe at high altitude consisting of a pressure of 1000 mb in January.  The top left shows the mean longitudinal velocity component, the top right shows the variance of the logitudinal velocity component, the bottom left shows the mean latitudinal velocity component, and the bottom right shows the variance of the latitudinal velocity component.  We compute the statistics using the data from \cite{kalnay1996} as described in the text.   The scales go from purple to red from -21 to 17 m/s, 2.5 to 15 m/s, -8.8 to 8.7 m/s, and -0.86 to 8.3 m/s for the plots, respectively.}
\end{figure}

To determine the distribution of thermodynamic quantities at the surface, we use the National Centers for Environmental Protection/National Center for Atmospheric Research (NCEP/NCAR) database which provides extensive information across the globe from 1948 to the present day (\citealt{kalnay1996}).  For the surface temperature, surface pressure, and surface relative humidity, we use the entire database on 2.5x2.5 degree grid that makes a series of measurements 4 times per day.  For a given latitude/longitude position, we bin the measurements into monthly groups including all years, so we can capture the seasonal variation.  Then, within each monthly dataset we calculate four quantities:  the mean value, the standard deviation, the mean diurnal variation from the 4 daily points, and the standard deviation of the diurnal variation.   The diurnal variation mean and standard deviation are multiplied by a correction factor of due to the underestimate of the variation.  If the daily variation is only sampled 4 times and is approximately sinusoidal then the variation is underestimated by a factor of 1.11.  We repeat this calculation for the surface temperature, pressure, and relative humidity.  This then allows a representation of the seasonal and dirunal variation for every location on the Earth at a grid spacing of 2.5 degrees.  The results are shown in Figures~\ref{fig:label1},~\ref{fig:label2}, and~\ref{fig:label3}.

We also capture the wind vector patterns using another related data set from \cite{kalnay1996}.  We use the vector mean and standard deviation can be captured on a monthly basis since 1948.  However, here we can obtain different wind vectors at a series of different pressure locations which allows us to construct the velocity vector at different altitudes.  There are 17 altitudes corresponding to different pressure levels and therefore different altitudes.  We then compute the mean and standard deviation of each component of the velocity vector for each month of the year on each location on the Earth at a 2.5 degree grid.  The global wind patterns for two altitudes are shown in Figure~\ref{fig:label4} and~\ref{fig:label5}.

\subsection{Vertical Profiles of Thermodynamic Quantities}

\begin{figure}[htb]
\begin{center}
\includegraphics[width=0.99\columnwidth]{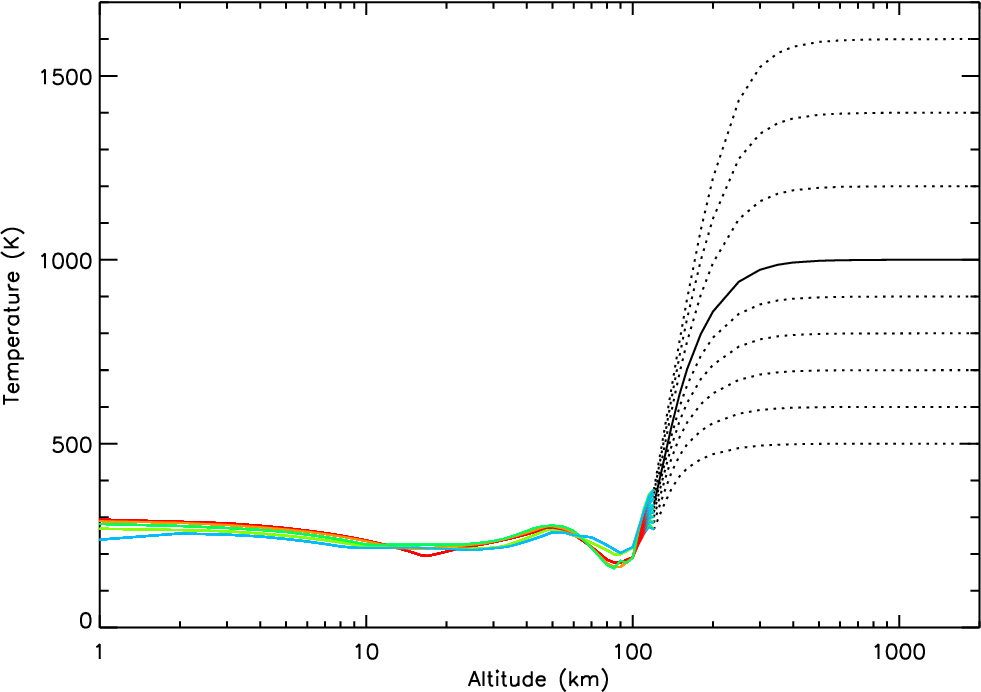}
\end{center}
\caption{\label{fig:label6} The temperature profile as a function of altitude.  The colored lines indicate the 5 different reference models (equator, mid-latitude winter/summer, sub-arctic from \cite{anderson1986}), and the profiles diverge at high temperature due to different asymptotic exosphere temperatures from \cite{jaccia1971}.  This interpolates and combines the results from \cite{anderson1986}, \cite{champion1985},  and \cite{jaccia1971}.}
\end{figure}

\begin{figure}[htb]
\begin{center}
\includegraphics[width=0.99\columnwidth]{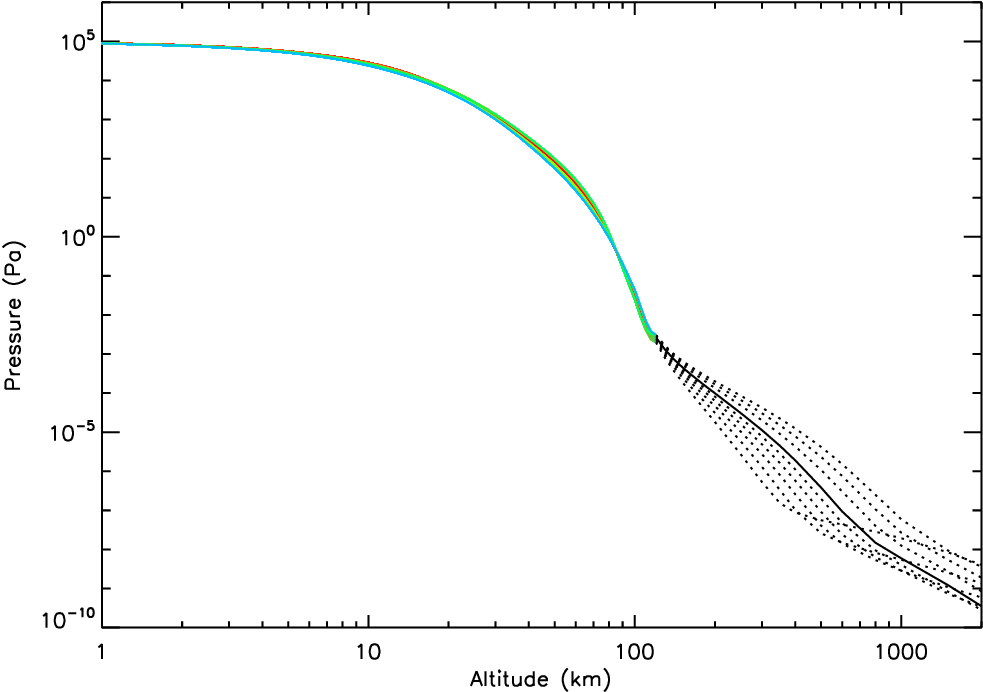}
\end{center}
\caption{\label{fig:label7} The pressure profile as a function of altitude.  The colored lines indicate the 5 different reference models (equator, mid-latitude winter/summer, sub-arctic from \cite{anderson1986}, and the profiles diverge at high temperature due to different asymptotic exosphere temperatures from \cite{jaccia1971}. This interpolates and combines the results from \cite{anderson1986}, \cite{champion1985},  and \cite{jaccia1971}.}
\end{figure}

\begin{figure}[htb]
\begin{center}
\includegraphics[width=0.99\columnwidth]{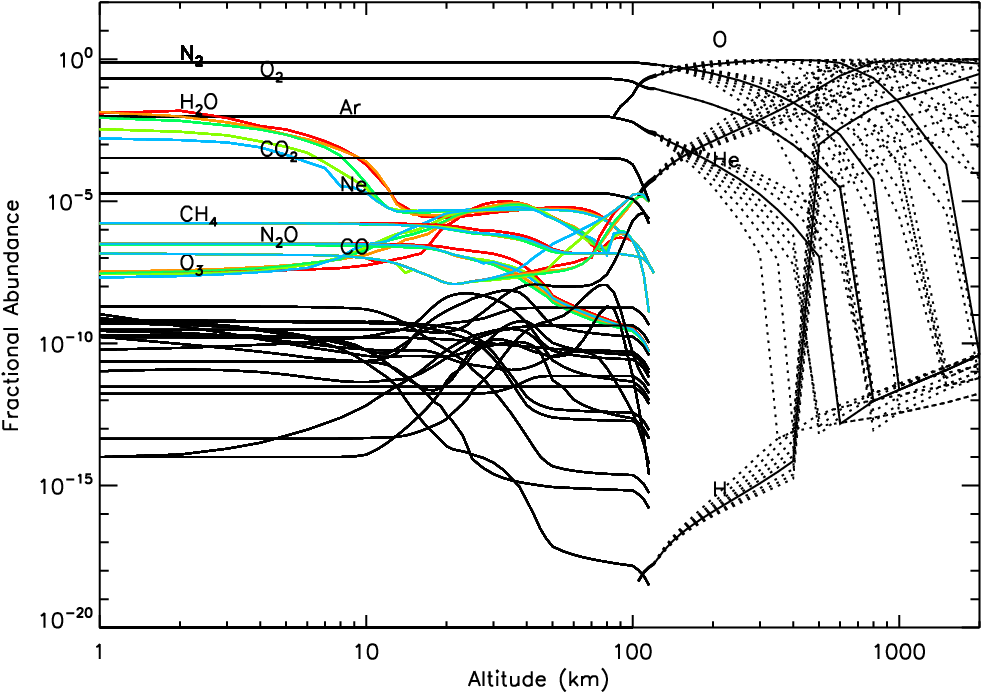}
\end{center}
\caption{\label{fig:label8} The relative abundance profiles as a function of altitude.  The most important molecular constituents are labelled.  The colored lines indicate the molecules that are most variable and are therefore different in the 5 reference models (equator, mid-latitude winter/summer, sub-arctic from \cite{anderson1986}).  The abundances diverge at high temperatures due to different exosphere temperatures from \cite{jaccia1971}.  This interpolates and combines the results from \cite{anderson1986}, \cite{champion1985},  and \cite{jaccia1971}.}
\end{figure}

The vertical profiles of thermodynamic and abundance quantities are not measured comprehensively for all sites at all times.  Consequently, they are produced from theoretical models that take into account the atmospheric chemistry (\citealt{smith1982}, \citealt{wmo1982}, \citealt{brasseur1984}, \citealt{wmo1986}). We use the Air Force Geophyiscal Laboratory (AFGL) constituent profiles (\citealt{anderson1986}) which comprehensively described thermodynamic and abundance profiles for different latitudes and seasons from ground level to 120 km.  The AFGL profiles have five generic profiles for the equator (tropical), a mid-latitude (45 degree) for winter and summer seasons, a sub-arctic (60 degrees) for winter and summer seasons, and a US standard.   The first 5 have pressure and temperature profiles and abundanced profiles for water, ozone, N$_2$O, CO, and CH$_4$.  The US standard profile includes the same quantities but adds many other molecules (CO$_2$, O$_2$, NO, SO$_2$, NO$_2$, NH$_3$, HNO$_3$, OH, HF, HCl, HI, ClO, OCS, H$_2$CO, HOCl, N$_2$, HCN, CH$_3$, H$_2$O$_2$, C$_2$H$_2$, C$_2$H$_6$, PH$_3$, Ar, O, Ne, Kr).
These additional molecules are not significantly time-variable unlike the other molecules (water, ozone, N$_2$O, CO, and CH$_4$).  For this reason, we can reasonably use the US standard abundance profile for the non-variable constituents regardless of the location or season.  Note for the last 4 elements above (Ar, O, Ne, Kr) we had to augment the data by assuming that Ar, Ne, and Kr assume their surface level value of \cite{champion1985} and the difference from the total is pure $O$, since the Noble gas species were left out of \cite{anderson1986} because of their negligible chemical interations.

Above 120 km, the chemical differences of the atmosphere are considerable.  Here we can use empirical-based models of \cite{jaccia1971}, which vary as a function of the exosphere asymptotic temperature.  The exosphere is relevant in our work, since it plays an important role in scattering sun light at twilight.  The profiles in \cite{jaccia1971} predict temperature and pressure profiles as well as molecular abundances of N$_2$, O$_2$, $O$, $Ar$, $He$, and $H$.  In \cite{jaccia1971} the exosphere ranges from 500 to 1600 K.  The exosphere temperature can either be specified by the user or we can estimate it based on the time of the observation.  Since the exosphere temperature is highly correlated with solar activity, we predict it using a sinusoidal function with the 11 year solar cycle period.  The mean value is set to 1050 K with sinusoidal amplitude of 550K.  We also linearly interpolate the previous values between 100 km and 120 km to asymptotically match the values of temperature, pressure, and abundances at 120km in order to avoid discontinuities.  The profiles combining the compilations are shown in Figures~\ref{fig:label6},~\ref{fig:label7}, and~\ref{fig:label8}.

For the temperature, pressure, and abundance profiles of the variable components, we use the following procedure to obtain realistic vertical profiles for any location for any time in the year.   Given the month and the latitude and longitude of the observation, we choose the appropriate mean, standard deviation, diurnal range, and diurnal range variation for the pressure, temperature, and relative humidity.  Then given the time of day of the observation, we then determine the surface pressure, temperature, and relative humidity by first gaussian sampling the mean value and the diurnal variation using the climate data in \S2.1.  Then we determine the temperature, pressure, and relative humidity by the following equations.

For the temperature variation, we use the \cite{gottsche2001}  model.  The temperature prior to sunset is given by

$$ T = T_0 + \left( T_1 - T_0 \right) \cos{ \left[ \frac{\pi}{\omega} \left( h - 14.2  \right) \right]}$$

\noindent The temperature after sunset is given by

$$ T = T_0 + \left(T_1 - T_0 \right) \cos{ \left[ \frac{\pi}{\omega} \left( T_s - 14.2\right) \right]} e^{-\frac{h-T_s}{k} }$$

\noindent where the time of Sunset is $T_s$ and the constant, $k$ is defined below.  We calculate the day length, $\omega$, as given by \cite{brock1981} as

$$ \omega = \frac{2}{15} \arccos{ \left( -\tan{\phi} \tan{\delta} \right)}$$

\noindent where $\delta$ is the Solar declination and is given by

$$\delta = 23.45 \sin{\left[ \frac{360}{365} ( 284 + d) \right]}$$

\noindent where $d$ is the current number of days in the year.  Then the constant, $k$, is defined by 

$$k = \frac{\omega}{\pi} \arctan{\frac{\pi}{\omega} (h - 14.2)}$$

\noindent
Similarly, the pressure, $P$, is defined by using a sinusoidal variation with a 12 hour period using

$$ P = P_{\mu} + P_{\sigma} \cos{ \left( 2 \pi \frac{ h - 10.5}{12} \right)}$$

\noindent and the relative humidity is defined using a full day sinusoidal variation

$$ h = h_{\mu} + h_{\sigma} \cos{ \left( 2 \pi \frac{h - 3.0}{24} \right) }$$

After selecting the temperature, relative humidity, and pressure, the profiles described above will be selected with some modifications.  We use a linear combination of the five profiles with weights according to:

$$ \frac{1}{(\frac{\Delta T}{10})^2 + \Delta \phi^2} $$

\noindent
where $\Delta T$ is the difference in temperature from the reference profile at ground level and $\Delta \phi$ is the difference in latitude from the reference profile.  This allows us to crudely sample a profile that would more closely resemble the location and season of the observation.  The appropriate linear combination of the temperature, pressure, and abundance profiles are calculated.   We temperature profile by a factor of

$$ \frac{T_0}{T_{ref}} \left( 1 - \frac{a}{a_{0}} \right) + \frac{a}{a_{0}}$$

\noindent
where $T_{ref}$ is the reference temperature at ground level of the five profiles, $T_0$ is the chosen temperature, $a$ is the altitude, and $a_0$ is a reference altitude set to 120 km.  The allows us to match the pressure, temperature, and water fraction at ground level perfectly and then asymptotically match the reference profile at high altitude.  This is a reasonable approximation to the weather conditions modifying lower layers of the atmosphere, but greater stability at higher altitudes.  We use an anologous factor for the pressure and water fraction as well (replacing the temperature above with the appropriate variable).  For the water fraction, we first calculate equivalent vapor saturation pressure by \cite{wagner1993} and \cite{wagner1994} using the selected temperature.  We then use the chosen relative humidity to predict a ground level abundance fraction using the selected pressure and the saturation pressure.  Finally, we use the same non-variable molecules abundance profile, but adjust the overall weight so the same proportion of non-water molecular ratios are obtained.  We then calculate the mean molecular weight as a function of altitude using values from \cite{pubchem}.  The density profile can be constructed assuming the ideal gas law, and using the temperature, pressure, and mean molecular weight profiles.

We also predict the potential temperature profile.  This is given by

$$ \theta = T \left( \frac{P_0}{P} \right)^{0.286} $$

\noindent
where $T$ is the temperature, $P$ is the pressure, and $P_0$ is the sea level pressure.  Since, potential is conserved for a parcel of fluid moving adiabatically , the derivative of potential temperature ($\frac{d \theta}{dz}$) can be use to predict the relative fraction of turbulence.  Specifically, we find

$$ \mu_{C_N^2} = 50.0 n k \frac{d \theta}{dz} $$

$$ \sigma_{C_N^2} = 2 \times 10^{-18} n \mu_{C_N^2}$$

\noindent
where $n$ is the number density.  The result is to be able to predict the temperature, pressure, humidity, and turbulence intensity as a function of altitude for any time and location on the globe while anchoring the results to the theoretical calculations in the literature.  As discussed above, these quantities can be input by the user, but the ability to predict reasonable values for any time and location will be useful.

\subsection{Cloud Representation}

\begin{figure}[htb]
\begin{center}
\includegraphics[width=0.99\columnwidth]{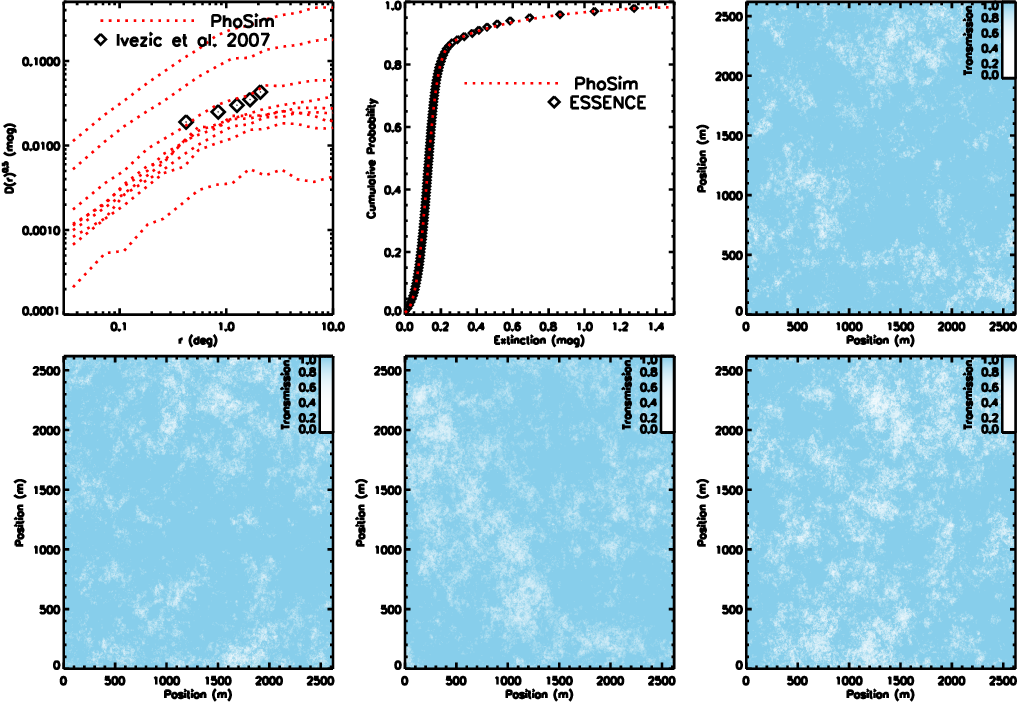}
\end{center}
\caption{\label{fig:label9} Cloud opacity simulations and comparisons.  The top left panel shows various realizations of the structure function from PhoSim compared with the single SDSS measurements of \cite{ivezic2007}.  The average of the PhoSim curves is similar.  The structure function normalization will change with various levels of opacity, but we match the average structure function.  The top middle panel shows the cumulative distribution comparison with Essence by \cite{miknaitis2007}.  The other panels show the variety of opacity patterns on 2 km scales setting the areas where there is no absorption to blue and high absorption in white.}
\end{figure}

There are a number of aspects of clouds to fully capture their effect on astronomical images.  Their spatial structure can be characterized by a series of frozen screens at various altitudes that drift during the exposure according to the local wind vector.  We first need to characterize the optical depth to extinction (scattering or absorption) at each point.  The morphology of clouds patterns and types is not well characterized in a mathematical representation in the literature.  Therefore, we first attempt to represent this in a complex manner to capture the major details for physics simulation.  We present the optical depth spatial position with a spatial power spectrum.  We find that a spatial power spectrum following the scale, $l^{\frac{1}{3}}$, in 1-dimension that reproduces spatial statistics described below.  This is flatter than the Kolmogorov spectrum that follows, $l^{\frac{5}{3}}$, commonly considered for atmospheric turbulence.  We then generate integrated screens having spatial power spectra following this spectrum.  The amplitude therefore has a unitless variance.  We then represent the optical depth pattern in proportion to the screen amplitude, however, an important physical aspect of clouds is that there are many lines of sight where the optical depth is negligible (i.e. they are scattered) that possibly represents a threshold thermodynamic quantities required for condensation.  To represent this, we make the magnitude of the optical depth, $m$, equal to

$$ m(x,y) = max( \mu + \sigma s(x,y) , 0)$$

\noindent
where $s(x,y)$ is the screen pattern, $\mu$ is the mean, $\sigma$ is the standard deviation.  The maximum plays the role of a ramp or rectifier function that then will make a significant fraction of the pattern of being cloud free.  The mean and standard deviation can then be predicted from historical extintion distributions.   The mean, which affects the cloud cover in the above formula, is also significantly affected by the relative humidity with considerable scatter.  Therefore, we choose the cloud cover, $c_c$ according to

$$c_c = \alpha e^{ -\frac{1-r}{\beta}} + \gamma g $$

\noindent
where $r$ is the relative humidity at ground level, $\alpha$ is the cloud cover at 100 percent humidity, $\beta$ sets the fall off with humidity, $\gamma$ is the amplitude of scattering using a Gaussian random number, $g$ which expands on the representation of \cite{walcek1994}.
This then will capture the cloud cover correlations with humidity and therefore diurnal and seasonal cycles represented above.   The other variable besides cloud cover will be the cloud depth, $c_d$, which we define as the average cloud extinction.  We choose this variable as a uniform distribution up to a maximum value, $c_d^M$.   We derive the mean, $\mu$ and standard deviation, $\sigma$ from a truncated distribution from the cloud cover, $c_c$ and the cloud depth, $c_d$, as

$$\sigma =\frac{c_d}{\sqrt{2} f c_c +\frac{1}{\sqrt{2 \pi} e^{-\sqrt{\pi} f}}}$$

$$\mu = \sqrt{2} \sigma f$$

\noindent
where $f$ is the inverse error function of $2 c_c -1$.  Thus, we can generate the screens, $\mu$, and $\sigma$ by from two variables, $c_c$ and $c_d$.  Those variables can either be specified for a given simulation or predicted randomly that also incorporates the humidity distributions.

To set the constants in the above equation, we use extinction distribution in the ESSENCE survey (\citealt{miknaitis2007}, \citealt{woodvasey2007}, \citealt{narayan2016}) where the photometry across a multi-year survey was carefully tracked in order to estimate accurate Supernovae photometry.  The extinction distribution of this survey has a narrow peak and a long tail where the clouds dominate as shown in Figure~\ref{fig:label9} (C. Stubbs, priv. comm.).   The narrow peak is consistent with the affect of aerosols having $0.12 \pm 0.05$ magnitudes of extinction that is similar to what is expected from the aerosols distributions described below for the Blanco site.  The tail of the distribution can then be used to predict the mean, $\mu$ and the standard deviation, $\sigma$ to be used in the screen calculation above.   Using the humidity distribution for the Blanco site, we predict the cloud cover fraction and random predict the cloud depth up to a maximum values.  We find $\alpha = 0.65$, $\beta = 0.3$, $\gamma=0.1$, $c_d^M = 0.5$, fits the cumulative distribution well as shown in Figure~\ref{fig:label9}.  The first three parameters are not strongly constrained by a single data set, but this is also consistent with studies of cloud cover in \cite{walcek1994} as well as the maximum global cloud cover being near 65\% for humid sites.

We then intergrate the screens by using the wind distributions for the screen height as described above for the wind distributions.  The height of cloud in the simulation is also a complex topic.  However, a simple prescription is to place the screens at the approximate height where the water would achieve 100\% humidity given the relative humidity at ground level.  Therefore, this is given approximately by $ 8.5 \mbox{km} \log{h}$ where $h$ is the relative humidity at ground level for a pressure scale height of 8.5 km.  We also split the cloud distributions at the two closest levels to give some verticality to the distribution and allow for different velocity vectors.  The cloud depth of an individual layer, $c_i$  has to be $c_i = 1 - \sqrt{1-c_t}$ to preserve the total cloud cover, $c_T$.    The cloud depth can be divided between the two layers equally.  Despite, the complexity the entire cloud representation, two parameters (the cloud cover and cloud depth) capture most of the observational characteristics.

We then consider the validation of the spatial statistics of the extinction variation.  The photometric variation of bright stars on small spatial scales using the SDSS observations was studied in detail by \cite{ivezic2007}.  Here we can match the spatial statistics which tests both the input power spectrum as well as the use of a ramp function described above.   Figure~\ref{fig:label9} shows the structure function of the \cite{ivezic2007} results as well as the structure function of the screens.  On average there is good agreement with the average structure function.  Future work may allow us to look at structure functions in a variety of conditions and sites as well as the average photometry statistics, but this representation is sufficient given the current set of measurements.  The interaction of light with the clouds is described in \S3.

\subsection{Aerosol Representation}

\begin{figure}[htb]
\begin{center}
\includegraphics[width=0.99\columnwidth]{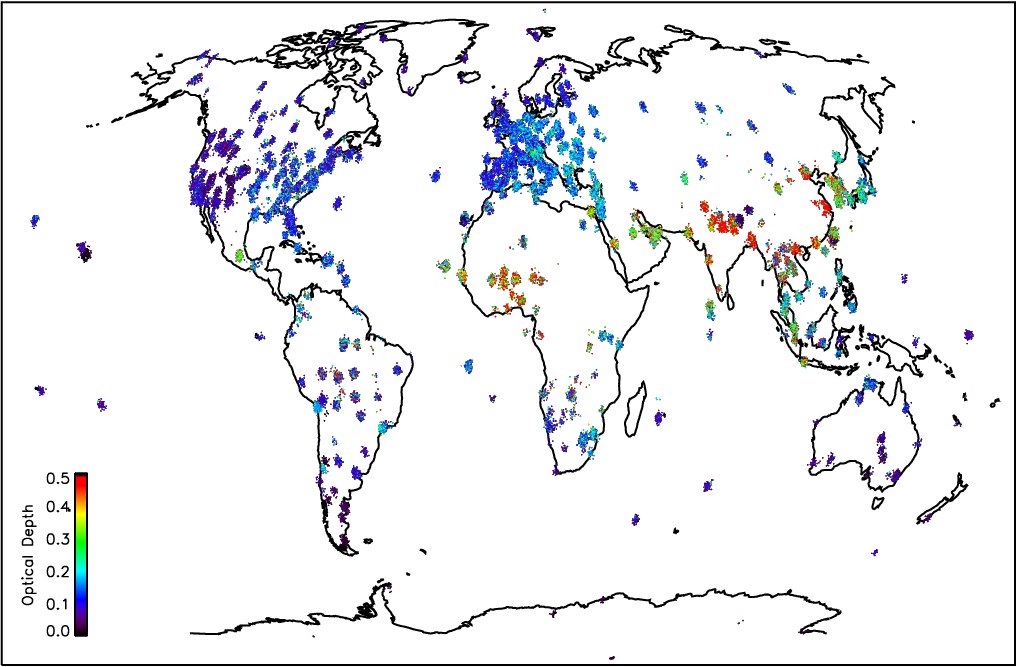}
\end{center}
\caption{\label{fig:label10} Global fit to the AERONET data (\citealt{holben1998}, \citealt{aeronet}) for various sites across the global.  The color corresponds with the optical depth where blue indicates negligible optical depth and red indicates optical depth of 0.5.}
\end{figure}
\begin{figure}[htb]
\begin{center}
\includegraphics[width=0.99\columnwidth]{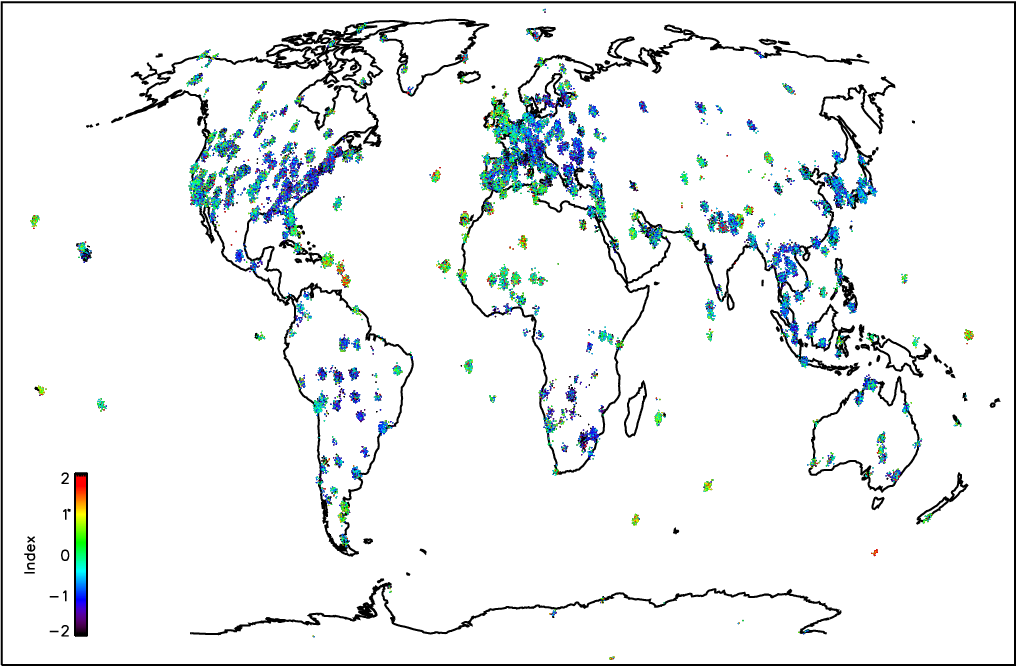}
\end{center}
\caption{\label{fig:label11} Global fit to the AERONET data (\citealt{holben1998}, \citealt{aeronet}) for various sites across the global.  The color corresponds with the spectral index where blue indicates an index of -2 and red indicates an index of 2.}
\end{figure}
\begin{figure}[htb]
\begin{center}
\includegraphics[width=0.99\columnwidth]{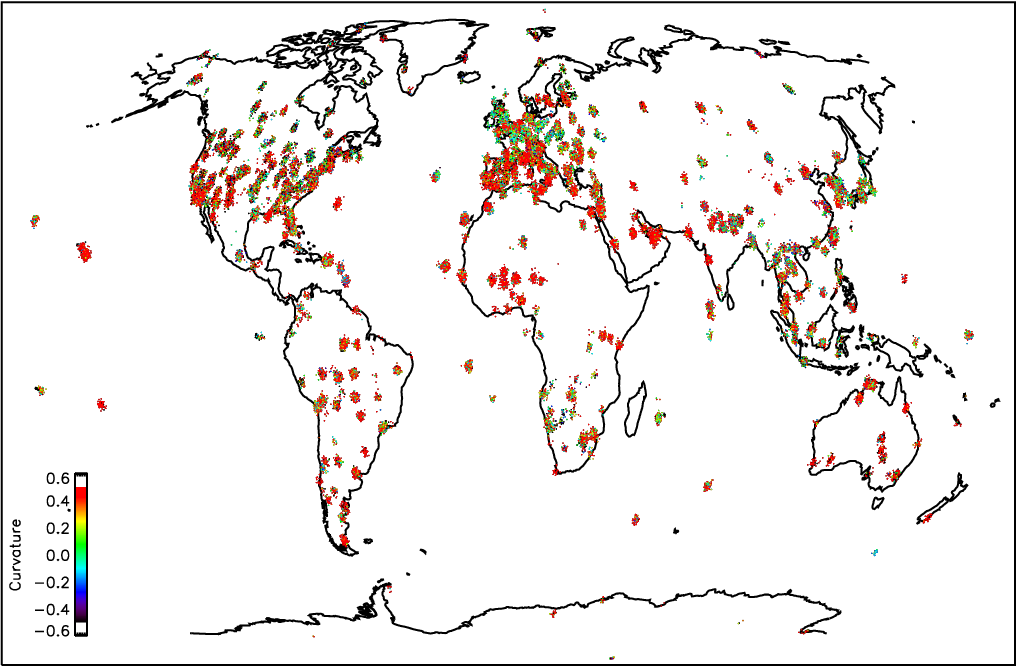}
\end{center}
\caption{\label{fig:label12} Global fit to the AERONET data (\citealt{holben1998}, \citealt{aeronet}) for various sites across the global.  The color corresponds with the spectral curvature where blue indicates a curvature of -0.5 and red indicates a curvature of 0.5.}
\end{figure}

The final component of the atmosphere is aerosols that arise from a number of sources.   The dominant cause of large particles are from dust, pollution, smoke, and sea salt.  There is a wide variation from site to site and there is likely a mixture at all locations.  In addition, there is a significant seasonal variation as well.  To address this comprehensively, we use the extensive database of AERONET sites that covers a series of measurements across the globe (\citealt{holben1998}, \citealt{aeronet}).  We use 1271 sites where monthly data is collected on aerosol optical depths as a function of wavelength for up to the last 25 years.  For each monthly measurement at a given site, we fit the optical depth as a function of wavelength to a power law with possible curvature as:

$$ \tau = e^{p_0+p_1 \log{\frac{\lambda}{500.0}} + p_2 \left( \log{\frac{\lambda}{500.0}} \right)^2}  $$

  \noindent
  where $\tau$ is the optical depth, $\lambda$ is the wavelength in $nm$, and $p_0$ , $p_1$, and $p_2$ are the normalization, index, and curvature of the power law.  Typically, at least a half dozen wavelength measurements are obtained so it is well-constrained.  The measurements are shown in Figure~\ref{fig:label10},~\ref{fig:label11}, and~\ref{fig:label12}.

  To separate out the aerosol components the following procedure is adopted.  We first found regions on the globe where each of the four components (dust, pollution, smoke, and sea salt) was the dominant aerosol.  We use the Sahara desert for dust, the Amazon and Congo rainforests for smoke, isolated islands far from continents for sea salt, and the largest population density areas for pollution.  For each case, the mean value of the three parameters can be obtained as well as the standard deviation.  In addition, there is sufficient data to crudely estimate the altitude dependence of each component by comparing different sites at various altitudes.  We use altitude for every site using the NOAA GLOBE (Global Land One-Kilometer Base Elevation) data set on 1 km scale (\citealt{globe}).  We use the population density for every point on the globe from the CIESIN data set (\citealt{ciesin}).  The altitude dependence has an exponential scale height of 2.2 km for seasalt, 3.6 km for dust, 6.6 km for smoke, and 1.7 km for pollution.

  Then, we need a method to estimate the relative contribution of various components at a given site.  There are a variety ways to do this with a detailed model, but a relatively simple prescription is to predict:  1) the sea salt component in proportion to the nearby land area, 2) the smoke component in proportion to the temperature and rainfall, 3) the dust component negatively in proportion to temperature and rainfall, and 4) the pollution component in proportion to the population density.  The rainfall is predicted by computing the average daily water vapor pressure times the humidity divided by the density of water and gravitational acceleration.   For a given site, we consider the site and four points 1000 km away in each cardinal direction to form a weighted sum.

 $$ S = \sum_i \left( 1 - L_i \right)  w_i e^{-2.56  +  0.29 g}$$
 $$ D = \sum_i L_i \max[ \left( -r_i + 0.2 t_i \right), 0]  \frac{1}{0.12} e^{-0.8 + 0.46 g}$$
$$  S_m =\sum_i L_i w_i \max[ \left( r_i - 0.2 t_i \right), 0] \frac{1}{0.06} e^{-1.83 + 0.8 g }$$
 $$ P=\sum_i w_i p_i \frac{1}{0.14} e^{-1.28 + 0.4 g}$$

  \noindent
  
  where $S$ is the seasalt, $D$ is the dust, $S_m$ is the smoke, $P$ is the pollution, $p_i$ is the scaled population density, $t_i$ is the scaled temperature, $r_i$ is the scaled rainfall, $L_i$ is a function that is 1 when the location is above sealevel and 0 otherwise, and $w_i$ is the weighting function.

  Finally, we need to determine the aerosol particle sizes in order to perform Mie scattering in \S4.  To do this, we take the spectral index and curvature determined for each of the four components.  Then we consider particles having a log normal distribution with a mean and variance.  For each component, using Mie scattering in \S4, we choose combinations of mean and variance that are consistent with the range of spectral indices and curvature for the extinction.  We also consider water droplets for the clouds and assume that the extinction is approximately flat.  We use the index of refraction for the dominant molecular for each component: water (clouds), sodium chloride (sea salt), carbon (smoke), silica (dust), and ammoniun sulfate (pollution).  In general, there are a range of particle size distributions that are consistent with the extinction and may represents a combination of the true possible distributions as well as some degeneracy due to the finite information in the extinction spectrum.  We find the following parameters to choose the log-normal distributions with a correlation between the according to:

  $$ \mu =  p_1 + p_2 g_1$$
  $$ \sigma = p_3 + p_4 g_2 + (\mu - p_1) p_5$$

  \noindent
  where $p_1$ and $p_2$ set the mean particle size in microns, $p_3$ and $p_4$ set the variation, $p_5$ induces a correlation, and $g_1$ and $g_2$ are gaussian distributed random numbers.  The particle distribution is then chosen according to $ e^{\mu + g \log \sigma}  $, where $g$ is a gaussian random number chosen for each particle.  The parameters are given in the Table~\ref{tab:table1}.  Alternatively, the aerosol optical depth for the four components can be specified by the user for a particular location and observation time, but the distributions by the calculation above should produce reasonable values for any location and observation time.

\begin{deluxetable}{lrrrrr}
  \tablecolumns{6}
  \tablecaption{\label{tab:table1} Relationship between important environmental variables and image characteristics }
  \tablehead{
    \colhead{Type} &  \colhead{$p_1$} & \colhead{$p_2$} & \colhead{$p_3$} & \colhead{$p_4$} & \colhead{$p_5$}
    }
  \startdata
  Water &  0.7 & 2.6 & 0.93 & 0.40 & -0.08 \\
  Sea Salt & -4.2 & 1.3 & 1.13 & 0.11 & -0.20 \\
  Dust & -3.6 & 1.2  & 1.16  & 0.10 & -0.22 \\
  Pollution & -4.3 & 1.3 & 0.99 & 0.04 & -0.21 \\
  Smoke & -4.9 & 1.2 & 0.95 & 0.06 & -0.22 \\
  \enddata
  \end{deluxetable}

\section{Photon Interactions}

The simulation of photons proceeds by following the photon through layers of the atmosphere in a forward raytracing scheme (\cite{peterson2015}).  This is implemented using first principle physics and acts on the atmosphere representation described in \S2.  In each layer, we consider 1) the refraction, 2) the interaction with the layer of air through either Rayleigh or molecular scattering and absoprtion, and 3) the possible interaction with the aerosols in the bulk layer and cloud droplets in the cloud screens by Tyndall-Mie theory.  Following, we discuss each of these photon interactions.  When dealing with background light, a modified procedure that is more computationally efficient is used, that is discussed separately below.

\subsection{Refraction}

The index of refraction in air, $n$, is given by the simplified formula of \cite{owens1967}

$$ (n -1) 10^{8} =  (2371.34 + \frac{683939.7}{130 - \sigma^2} + \frac{4547.3}{38.9 - \sigma^2} ) D_s$$
$$+ (6487.31 + 58.058 \sigma^2 - 0.71150 \sigma^4 + 0.08851 \sigma^6) D_w$$

$$ D_s = \frac{P_s}{T} \left[ 1 + P_s \left( 57.9 10^{-8} - \frac{9.3250 10^{-4}}{T} + \frac{0.25844}{T^2} \right) \right] $$

$$ D_w = \frac{P_w}{T} \left[ 1 + P_w \left( 1 + 3.7e-4 P_w \right) \left( -2.37321 10^{-3} + \frac{2.23366}{T} - \frac{710.792}{T^2} + \frac{7.75141 10^{4}}{T^3} \right) \right] $$

\noindent
where $T$ is the temperature in Kelvin, $P_s$ is dry air pressure, $P_w$ is the water pressure, and $\sigma$ is the inverse wavelength.  Light will be refracted throughout its trajectory as it passes through the atmosphere.  It is important to note that if the Earth were flat, then only the index of refraction near the surface would matter.  This is because plane-parallel applications of Snell's law essentially would cancel and only the final refraction step would be relevant (\citealt{filippenko1982}).  Since, the Earth is curved this simplification is not valid, and the complete layer by layer approach is necessary taking into account the full geometry (\cite{stone1996}).  We extend this approach in three important ways by using a numerical Monte Carlo integration approach.  One is that we have a range of temperature, pressure, and water vapor profiles which are naturally that are handled naturally by these methods.  This allows us to assess the variation of atmospheric dispersion under different conditions.  Second, we can more accurately predict dispersion particularly at large zenith angles, since other analytic approaches involve expansions in tangent of the zenith angle which are known to be catastrophically inaccurate at large zenith angles.  Finally, we can also perform dispersion for a individual photon at a time, which allows us to accurately predict the range of emergent effects for non-monochromatic and spectrally-varying spatialy-extended sources as we highlight in \S4.1.

The atmospheric dispsersion is handled by first considering the direction to the zenith position in the local coordinate system.    We use the fact that the refraction invariant, $r_0$, remains constant throughout the refraction process (\citealt{fabritius1878}, \citealt{young2004}).

$$ r_0 = (h + R_E) n \sin{z}$$

where $R_E$ is the radius of the Earth, $h$ is the height above sea level, and $z$ is the zenith angle.  The geometry can be written in differential form so that the change in angle, $\Delta \phi$ at any layer due to refraction is given by

$$ \Delta \phi = - \tan{z} \frac{\Delta n}{n}$$

\noindent
where $\Delta n$ is the change in index of refraction, $n$ is the local index of refraction, and $z$ is the zenith angle (\citealt{young2004}).  Before the photon enters the atmosphere at height, $h_0$, the refraction invariant is simply, $r_0 = (R_e + h_0) \sin(z_0)$, and $\tan{z}$ can be substituted as

$$ \Delta \phi = -\frac{(R_e + h) \sin{z_0}}{\sqrt{{\left[ n (R_e + h) \right]}^2 + {\left[ (R_e + h_0) \sin(z_0) \right]}^2}} $$

  \noindent
  Then the shift is performed in the direction of the zenith at each layer.  During the process both the index of refraction and the current height is altered, so that the approximately curved trajectory is created through numerical integration.  We also allow for the direction of the angle to the zenith to be modified during the exposure due to the field rotation, so depending on the arrival time of the photon the refraction may be different.  In many applications, we also allow an option to simultaneously subtract off a nominal refraction for a photon at the center of the field of view at a nominal reference wavelength for the filter, so that the overall nominal dispersion is cancelled as is customary for telescope operations.  We describe the effect of the dispersion in \S4.1.

\subsection{Rayleigh Scattering}

The Rayleigh scattering cross-section can be written as

$$ \sigma = \frac{24 \pi^3 \left( n^2 -1 \right)^2}{\lambda^4 N^2 \left( n^2 +2 \right)^2} F$$

where $n$ is the index of refraction (a function of wavelength and thermodynamic variables), $N$ is the number density, $\lambda$ is the wavelength, and $F$ is the polarization factor (e.g. \cite{hulst1957}).  We use the index of refraction described in \S3.1 and the density given the temperature, density, and water fraction at a given height when we evaluate this cross-section.  The calculation of Rayleigh scattering then resembles the \cite{tomasi2005} calculation but using \cite{owens1967} for the index of refraction.

\subsection{Molecular Interactions}

We simulate molecular interactions by using the HITRAN data base (\citealt{rothman2009}, \citealt{rothman2005}, \citealt{rothman2003},
\citealt{rothman1998}, \citealt{rothman1992}, \citealt{rothman1987}, \citealt{hitran}).  We calculate the cross-sections for molecular transmissions as described in \cite{peterson2015}.  In our new comprehensive representation, however, the molecular interactions lead to absorption and re-emission rather than simply removing photons (greenhouse effect).  This is important for the simulation of the scattered background light.

\subsection{Tyndall-Mie Scattering}

The aerosols and cloud droplets are large enough to require a full electromagnetic treatment.  Assuming the particles are quasi-spherical, the interactions with photons can be described by Tyndall-Mie theory.  This performs a self-consistent electromagnetic calculation, and predicts the absorption and scattering probability as a function of angle.  The result mainly depends on the dimensionless ratio, $ x = \frac{2 \pi r}{\lambda}$, where $r$ is the size of the particle and $\lambda$ is the photon wavelength.  It also depends on the real and imaginary part of the index of refraction as a function of wavelength.  For water, we use the optical constants of \cite{hale1973} for water,  Sodium Chloride constants of \cite{li1976} for seasalt, Silicon constants of \cite{aspnes1983} for dust, Ammonium Sulfate constants of \cite{cotterell2017} for pollution, and Carbon constants of \cite{phillip1964} for smoke.

Numerically, we perform the Mie calculation for two-dimensional grid of values of $x$ and $\lambda$.  We use the Bohren-Huffman algorithm originally described in \cite{bohren1983}.  This solves the electromagnetic boundary conditions iteratively, and results in a scattering angular probability distribution and absoprtion probability.  We then save these probability distributions and the total effective cross-section of both scattering and absorption.

In order to properly simulate the photon interactions with these particles, we need to construct a three-dimensional slab and perform photon Monte Carlo techniques to handle multiple scattering properly.  To do this, we simulate a photon incident on the slab and consider its three-dimensional trajectory.  We then choose a particle from the log-normal distribution for the aerosol type or cloud droplet distributions.  We calculate the mean free path given that particle size as

$$\frac{f^2}{\tau \sigma^2} \frac{1}{2}$$

  \noindent
  where $\tau$ is the optical depth, $\sigma$ is the particle size, the factor $\frac{1}{2}$ ensures that we allow for a cross-section four times more than the geometric area, and $f$ is a numerical constant that we fix depending on the log-normal distribution to achieve the correct optical depth.  $f$ can be determined by making

  $$ 1 - e^{-\frac{\tau \sigma^2}{f^2}} $$

  \noindent
  match $1 - \exp{-\tau}$ when the full log-normal distribution is considered for $\sigma$.  Then, the photon is shifted by a distance equal to the mean free path following an exponential distribution.

  Applying the photons angular trajectory, a new position in the slab is calculated.  Then, photons are removed if they are at a new position above the slab and are therefore reflected.   We roll a uniform random number to allow for the possibility that the photons are neither scattered or absorbed since we are assuming that the particle size are a factor of 4 above there geometric area, and for most cases the effective cross-section is near a factor of 2 more than the geometric area.  If they are still in the slab, then there is some probability of photon absorption according to the calculated absorption cross-sections.    If they are not absorbed then, they will be scattered.  We then use the cumulative probability distributions given from the Mie probability tables calculated earlier.  These are formulated in terms of the cosine of the polar angle, so the azimuthal angle is then chosen.  Then, the three-dimensional position is updated based on the new trajectory.  Finally, the process is repeated until the photon either emerges from the slab (either forward or backwards) or is absorbed.  Then, we will have a new perturbed trajectory for photons that have successfully scattered off of an aerosol particle or cloud droplet that is added to the existing trajectory before interactiion with the layer.  This process correctly handles multiple scatters and matches the local optical depth which can vary considerably in the case of clouds.

\subsection{Background Methodology}

\begin{figure}[htb]
\begin{center}
\includegraphics[width=0.99\columnwidth]{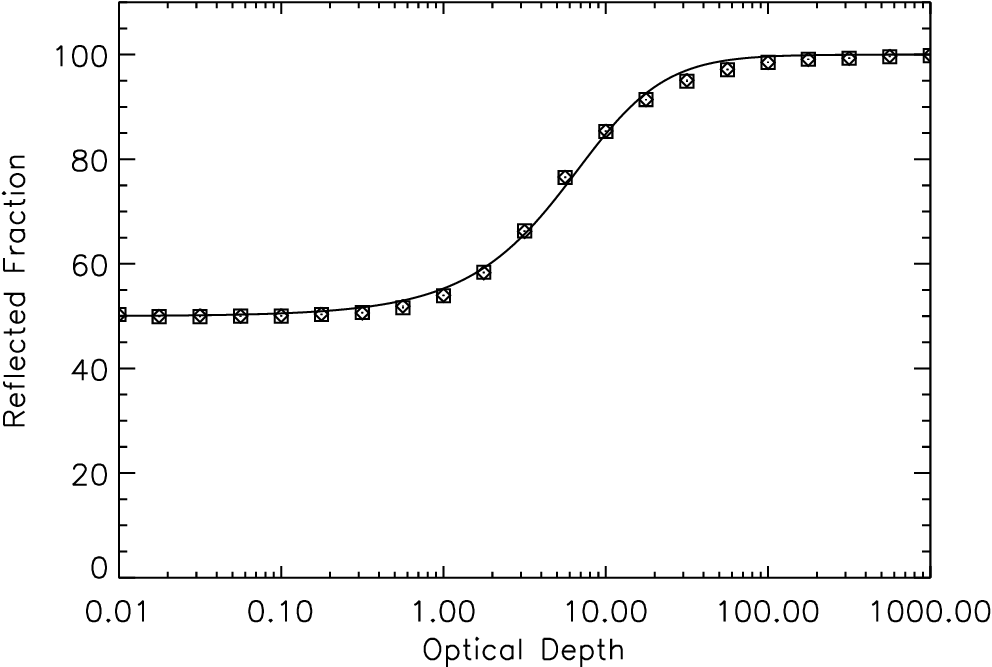}
\end{center}
\caption{\label{fig:label13} The calculation of the reflection fraction as a function of optical depth from a series of Monte Carlo simulations.   This is a Monte Carlo simulation of a generic slab of material to predict the amount of light that would reach the ground as a function of optical depth.  At high optical depth all light is reflected, whereas at low optical depth half is reflected.  The empirical fitting function is described in the text.}
\end{figure}

The simulation of the background is identical to the simulation of normal sources with some important exceptions to improve efficiency.  We start with background light from the Sun, reflected sunlight off the Moon, airglow with spatial variation, zodiacal light, and artificial light.  We then modify the simulation of light from that of normal sources to avoid simulating all background at all angles with possible scattering before deciding which photons would make it into a telescope aperture, which would be computationally prohibitive.  Essentially, we attempt to predict light from which angles would make it into the the telescope aperture before any significant computation.  This would be trivial, if there was no scattering, but scattering means the processes in the atmosphere have to be handled with care.  Essentially, this involves backward raytracing from the pupil plane.   This method is well known from computer graphics raytracing applications where most of the interest is in raytracing prior to the pupil plane (\citealt{goldstein1971}).  With the specific scattering in the atmosphere, we can perform this efficiently with the following methods.

The first exception is for direct backgrounds at the top of the atmosphere that includes zodiacal light and airglow.  These background are approximately flat, but have some small-scale spatial variation that is built into the simulator by having a spatially-dependent emission pattern.  Some fraction of this light can be forward Mie scattered by the methodology described above and would therefore be not detected in the field of view.  However, a nearly identical fraction of the light initially at angles outside the field of view would scatter back in the field of view, and the net effect would be simply to slightly blur the original spatial pattern of the emission.  Therefore, to simulate this we simply ignore forward Mie scattering for backgrounds of this type to a very good approximation.

A second related consideration is to properly handle Rayleigh and molecular scattering for these quasi-flat backgrounds.  These processes scatter light to much larger angles, and therefore remove a significant fraction of light when the optical depth is high for a given wavelength.  This is a straight-forward numerical calculation and is simply dependent on the optical depth.  From Monte Carlo simulations, we find the fraction of light that is forward scattered, if there is at least one scatter, as a function of optical depth is given by:

$$ 0.5 + \frac{5.28 \tau + 0.62 \tau^2}{1 + \frac{5.28 \tau + 0.62 \tau^2}{0.5}}$$

  \noindent
  where $\tau$ is the optical depth.  We performed Monte Carlo simulations for both an isotropic scatterer (i.e. molecular transitions) as well as Rayleigh scattering, and both were approximately the same.  Therefore, at low optical depth approximately half of the light that would continue to be scattered to the ground if there is at least one scatter, whereas at high optical depth no light would reach the ground.  Therefore, we use this function to reduce the opacity of the background light according to the optical depth, and therefore properly predicts the total light fraction.  The results are plotted in Figure~\ref{fig:label13}.

  A third consideration is to handle the scattered Moonlight and Sunlight.  In these cases, the scattered light itself is the main component of the background.  To simulate this light, consider an obervation at a point on the sky at some particular angle.   That angle will define a line segment from the telescope aperture to the top of the atmosphere.    Then, consider that a light ray from the Sun or Moon will intersect that segment uniquely.  Then, there will be a particular angle that forms by the intersection of  those line segments.  Therefore, in order for light to scatter it would need to scatter by this particular angle.  Therefore, to simulate this process we divide the observing segment into a set of smaller segments corresponding to layers of the atmosphere.  At each segment we consider the probability that light could scatter from a given process (Mie, Rayleigh, molecular).
  Then, if there is a point where we obtain a successful scatter, we consider removing the photon by attenuation prior to the scatter using a Monte Carlo approach.  Then, we continue the photon through the observing segment until it reaches the atmosphere.  It also could be scattered along this final path, so to properly handle this we simply require that the photon only scattered once.  This complex propagation scheme where we essentially are stepping though two partial line segments takes the same amount of computational time as a normal photon.  At the same time, we can treat all kinds of scattering properly and deal with a complex altitude dependence of scatterers including clouds and aerosols.

A final consideration is the simulation of artificial light (light pollution).  To simulate this, we simply simulate the light as if it came from the top of the atmosphere.  Then, we remove photons if there was not a scatter during the propagation.  This is equivalent to the correct physical simulation of light starting at the bottom of the atmosphere and only considering photons where they scattered backwards.  In all of these four cases, we are basically avoiding simulating any photons that do not have a significant chance of reach the pupil and then performing the appropriate physical scattering along the paths.

\section{Observational Effects}

There are a number of emergent effects that result from the photon interactions that are imprinted on the images.  Following, we discuss the effect on astrometry, point spread ellipticity, photometry, point spread size, and background.

\subsection{Astrometry and PSF Ellipticity}

\begin{figure}[htb]
\begin{center}
\includegraphics[width=0.99\columnwidth]{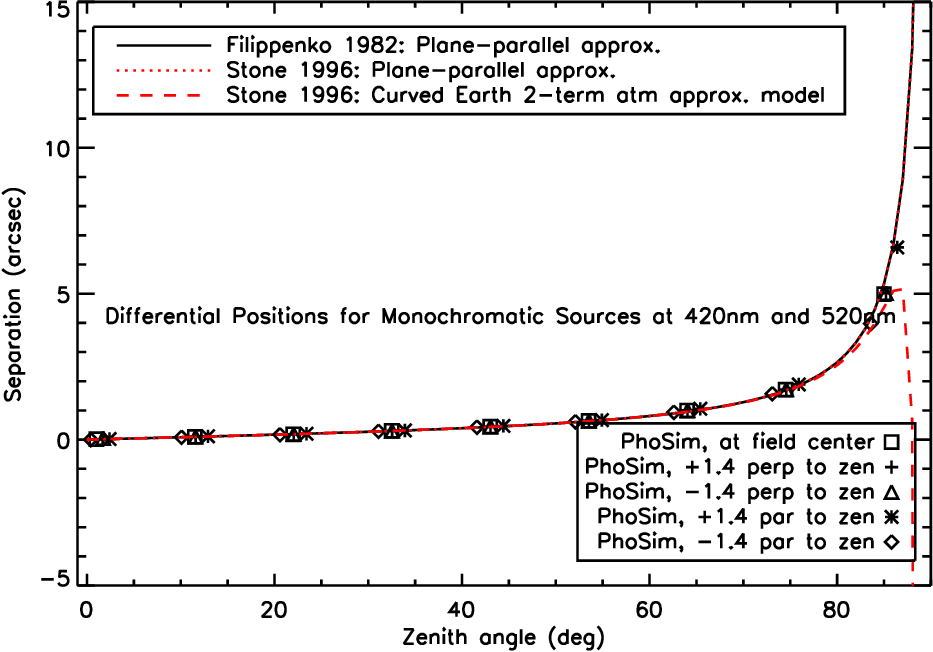}
\end{center}
\caption{\label{fig:label14} Differential chromatic refraction measured by the astrometric positions of two monochromatic sources of 420 and 520 nm as a function of zenith angle.  We include points at various offsets from the field center by 1.4 degrees from a generic telescope to demonstrate that the refraction works locally within a field.  The points are compared to three analytic calculations with good agreement at high altitudes (zenith angles less than 80 degrees).  Near the horizon the two plane parallel approximations fail since they use expansions in the tangent to the zenith angle.  The curved Earth two term model of \cite{stone1996} has better agreement at high zenith angle, but still diverges at high angles.  PhoSim can work to arbitrarily low angles since it does not involve any approximations, and can simulation telescopes at various altitudes.}
\end{figure}

There are several aspects of refraction that affect the astrometry of the field and ellipticity of effective point spread function.  First, the overall field is shifted by a significant amount as compared to the apparent position if the atmosphere was not present.  Second, there is a relative astrometric shift of some objects with respect to other objects depending on the relative position with respect to the zenith and the average wavelength of the detected photon (i.e. differential chromatic refraction).  Third, the detected photons with shorter wavelength will be shifted more to the zenith direction than the longer wavelength photons resulting in an elliptical shape.    A fourth effect is present when the astronomical source has a spatial distribution that varies significantly based on wavelength (i.e. has a color gradient).  This effect then results in a morphological chromatic differential refraction where the final distorted image is not simply described by the distortion by an elliptical point spread function as the refraction affects the morphological pattern in a wavelength-dependent manner.

The effect of differential chromatic refraction  is shown in Figure~\ref{fig:label14}.  Here we compare the predicted chromatic refraction with analytic models in the literature.
All four effects are essentially emergent effects due to the single physical effect of the refraction of a photon depending on the wavelength as it propagates through the atmosphere.  Therefore, in a photon Monte Carlo approach it is straight-forward to predict the complex emergent behavior.   In order to predict the lower altitude effect correctly, it is critical to pursue a layer by layer approach including the correct geometry of the curvature of the Earth otherwise the layer by layer refraction would cancel and it would predict infinite refraction at the horizon.

\subsection{Photometry}

\begin{figure}[htb]
\begin{center}
\includegraphics[width=0.99\columnwidth]{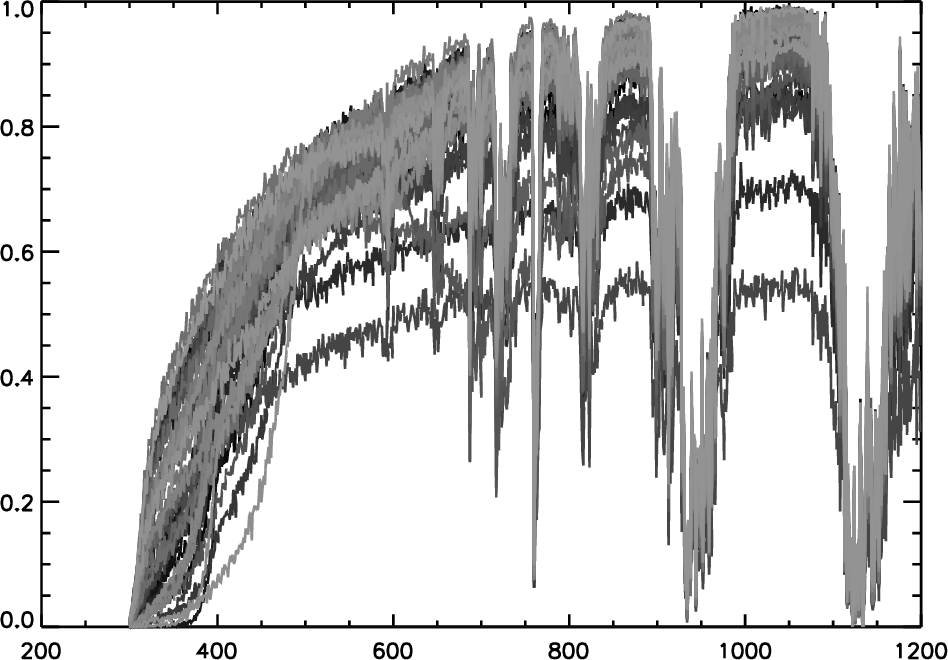}
\end{center}
\caption{\label{fig:label15} Examples of absorption profiles as a function of wavelength in nm at zenith.  The variation is significant the variation of different components (e.g water vapor, dust, clouds) including the height profile.  The different transmission functions are from completely different conditions (i.e. different day).}
\end{figure}

\begin{figure}[htb]
\begin{center}
\includegraphics[width=0.99\columnwidth]{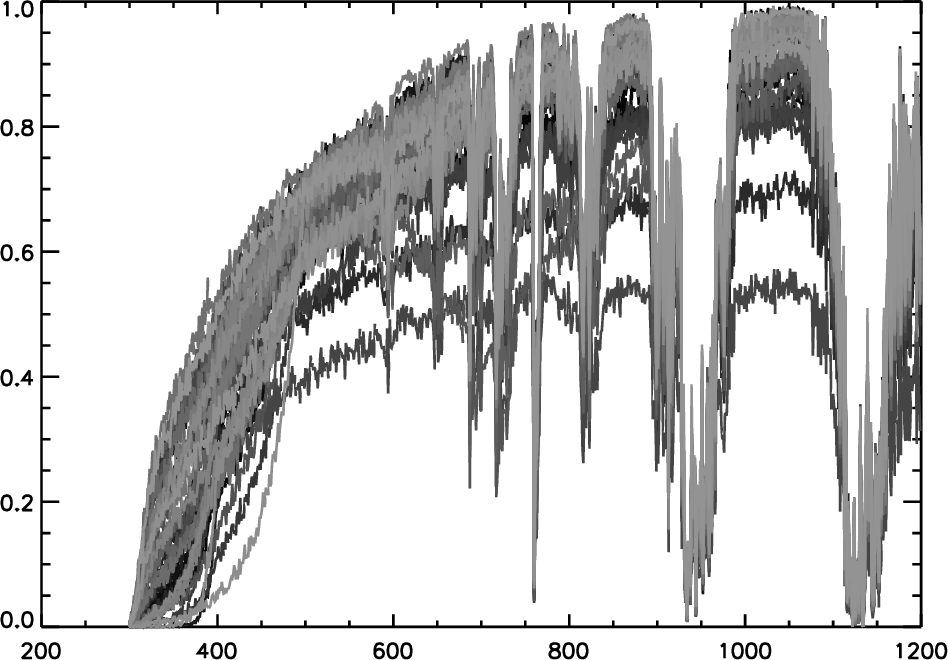}
\end{center}
\caption{\label{fig:label16} Examples of absorption profiles as a function of wavelength in nm at a zenith angle of 30 degrees.  The variation is significant the variation of different components (e.g water vapor, dust, clouds) including the height profile.  The different transmission functions are from completely different conditions (i.e. different day).}
\end{figure}

Light is attentuated by essentially all of the physical photon interactions described above.  Rayleigh scattering decreases short wavelength light, atomic and molecular absorption impacts specific frequencies, and Mie scattering and absorption decreases transmission depending on wavelength, particle size, and particle type.  The calculation correctly calculates a local path length (i.e. a local airmass), including the effect of the curvature of the Earth.  Mie scattering has a stronger wavelength dependence with smaller particles, so the pollution, smoke, and seasalt distributions attenuate blue light significantly.  Dust has a mild wavelength dependence, whereas the wavelength dependence of water droplets is minimal.

The atmosphere transmission is highly time variable, but this is predictable from the self-consistent atmosphere representation pursued in this work.  Figure~\ref{fig:label15} and~\ref{fig:label16} show a range of transmission spectra in the optical band for a zenith angle of 0 degrees and 30 degrees.  Significant time variation is present due to the variation of thermodynamic quantities (temperature, density), atmosphere molecular constituents (especially ozone, water), variation in cloud cover and depth, and variation in aerosols.  The cloud cover variation is clearly visible in a significant overall variation in the ceiling of transmission functions.  Interestingly, the aerosol variation is rather significant in the short wavelength part of the spectrum and points to a larger calibration challenge to u-band observations.  Standard techniques of adjusting zeropoints to match the photometry of bright calibration stars can clearly correct for some of this variation.  However, we can estimate the possible photometric variation that is not easily correctable.  If we simulate two stars separated by 3 arcminutes the average relative photometric variation is 20 millimags/$\sqrt{1 + \frac{t_{exp}}{2 s}}$. This would mean observations with short exposures (between 10 seconds and a minute) could be limited to  photometric accuracy typically to $0.3$ to $0.8\%$ with large-scale empirical calibration methods.   Only more elaborate physically-based interpolations that  essential map cloud patterns and molecular spatial gradients could achieve higher photometric accuracy.

\subsection{Point-Spread-Function Wing}

\begin{figure}[htb]
\begin{center}
\includegraphics[width=0.99\columnwidth]{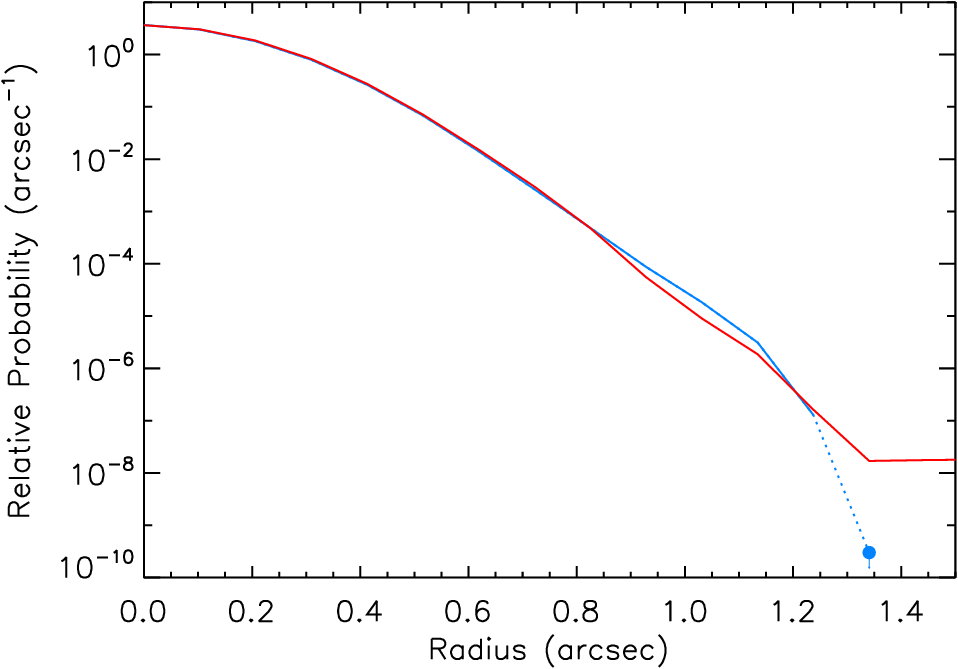}
\end{center}
\caption{\label{fig:label17} Average atmosphere point spread function for 200 different atmospheres mostly due to atmospheric turbulence.  The red curve shows the effect of Mie scattering of clouds whereas the blue curve removes the scattering.  The PSF wing starts to become significant beyond 1 arcsecond if other sources of the PSF wing (such as mirror microroughness) are negligible.}
\end{figure}

The Mie scattering from both clouds and aerosols results in considerable forward (small-angle) scattering.   This is shown in Figure~\ref{fig:label17}.  If this scattered light is at sufficiently small angles (i.e arcminutes), it will appear as a wing on the point spread function.  The scattering of particles happens at all angles according to the Mie formalism, but some fraction of the light will scatter at smaller angles.  This will depend primarily on the ratio of particle size to photon wavelength.  The forward scattering happens disproportionately for larger particles for a given wavelength in the atmosphere, since the smaller particles asymptotically approaches the Rayleigh limit (which has a modest  $1+\cos{\theta}^2$ dependence) for a given wavelength.  Similarly, the shorter wavelength light will be scattered more for a fixed particle size.

In our particle distribution for clouds and aerosols described above, the largest particles will be the water droplets.  Therefore, we the PSF wing is dominated by scattered light from clouds.  Similarly, the second largest contributor to the PSF wing is the dust aerosols that have a distributions that extend to larger sizes.  The seasalt, pollution, and smoke scattering results in a minimal distribution to the wing only from the largest particles.  In addition, for the same reason the scattering to small angles occurs disproportionately for shorter wavelength light since the Mie forward scattering occurs for larger ratios of the particle size over the wavelength.  For this reason, we predict a larger PSF wing in the UV part of the spectrum.  The wing from this scattering is the dominant contributor at larger angles (arcminute).  At smaller angles (several arcseconds), the PSF wing can have a significant contribution from turbulence, diffraction, or mirror microroughness.

Because the predicted PSF wing results primarily from water droplets in clouds, there will be a correlation with the overall extinction from clouds, since the larger extinction results in more scattering.  This is intuitively familiar as a {\it fog-lamp} effect where a car's headlight in fog is both attenuated and blurred (scattered) depending on the optical depth.~\footnote{Perhaps the best artistic visualization of this is the exaggerated haloes around V. Van Gogh's stars.}  Thus, we also expect considerable variation in the PSF wing, since the level of extinction can vary significantly in time as well as across the field.  There will then be an anti-correlation between an object's intensity and the PSF size (see \S4).

\subsection{Background}

\begin{figure}[htb]
\begin{center}
\includegraphics[width=0.99\columnwidth]{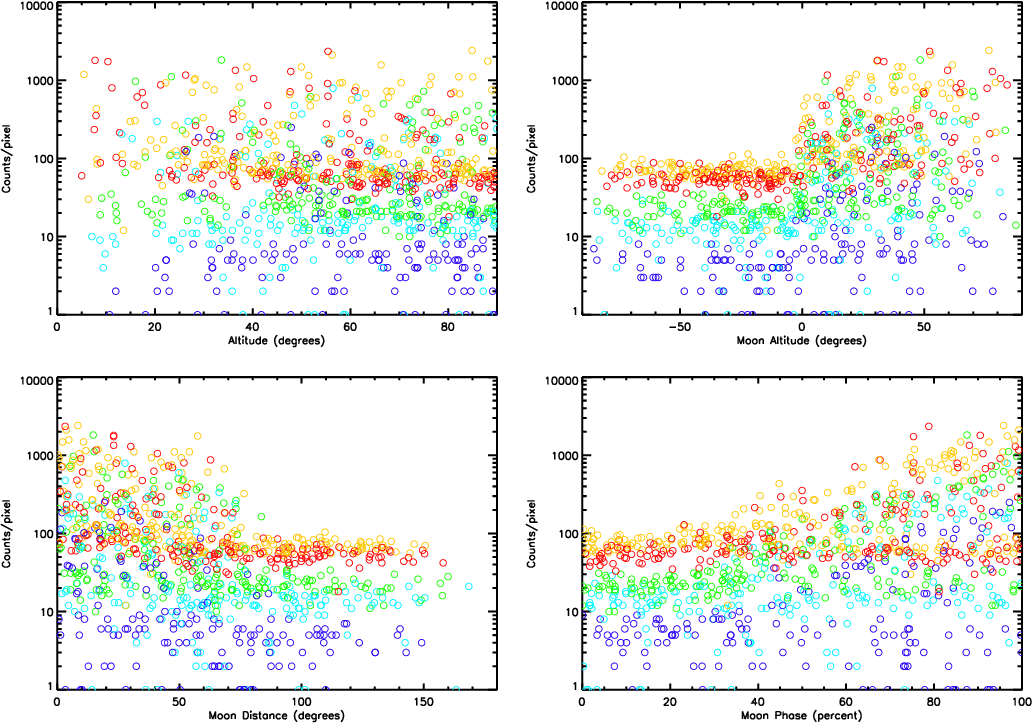}
\end{center}
\caption{\label{fig:label18} The simulated background level as a function of some of the more important variables:  altitude of the observation (top left), moon altitude (top right), distance to the Moon (bottom left) and the lunar phase (bottom right).   The different colors represent the u, g, r, i, and z intensities (blue, light blue, green orange, red, respectively).  The significant complexity is evident as the intensity varies by four orders of magnitude.  The simulation is for a 1 m f/4 telescope located at a latitude and longitude of (0,0) with 10 microns pixels and a exposure time of 15 seconds.  Results can be scaled for other telescopes and exposure times by multiplying by $\frac{t_{exp}}{15~s} {\left( \frac{p_{size}}{10~\mu m} \right) }^2 {\left( \frac{4}{f_\#} \right) }^2$.}
\end{figure}

\begin{figure}[htb]
\begin{center}
\includegraphics[width=0.99\columnwidth]{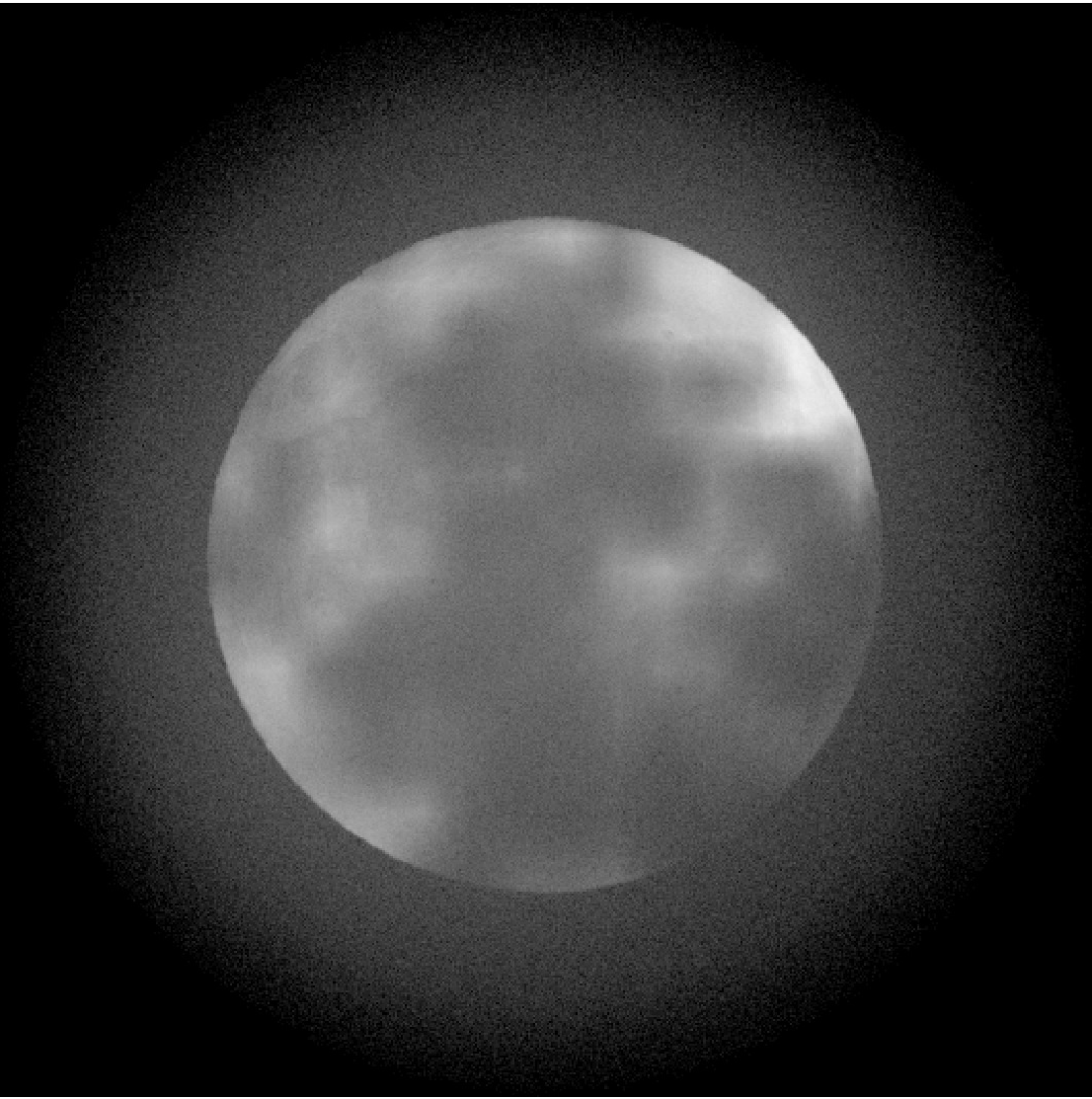}
\end{center}
\caption{\label{fig:label19} Simulation of the Moon with two observable effects.  The cloud opacity is visible across the Moon, and the Mie scattering halo is present near the edge of the Moon.}
\end{figure}

The overall background level is an emergent quantity that is predicted from a combination of physical processes described above.  Background, in particular, is related to two major effects:  1) the deviation from an ideal astrometry mapping patterns which distorts the incident background pattern and makes the pixels cover a varying solid angle, and 2) the wavelength-dependent photometry that has some spatial dependence in the direction of the horizon due to atmospheric transmission.  We also performed a simulation of the non-Lunar background on 10 by 10 arcminute scales, and found a variation in excess of the usual Poisson variation of 0.2\%.

The overall background level is a function of several variables.  The most important are the altitude of the observation, the Lunar phase, the altitude of the Moon, and the distance between the observation and the Moon.  The cloud patterns, aerosols, and the variation of atmospheric absorption lead to additional variation.  To assess this, we simulate 1000 random simulations where:  1) the observation's position is chosen randomly, 2) the Moon's position is chosen randomly (including below the horizon), 3) the Lunar phase is chosen randomly, 4) the filter is chosen randomly, and 5) all other variables (clouds, aerosols, atmospheric consituents) are chosen according to the distributions described above.  In Figure~\ref{fig:label18}, we demonstrate the variation in background by nearly 4 orders of magnitude.  Predictably, the background increases dramatically as the Moon is above the horizon and closer to the full Moon phase.  The background also increases dramatically at smaller lunar distances due to increase of Mie scattering at  smaller angles in comparison to Rayleigh scattering.  Interestingly, the non-Lunar background also shows a net increase at low altitudes due to the relative increase in emission volume from airglow.  This is only partly modified by atmospheric attenuation which is more visible in the bluer bands.

A demonstration of the complexity of the background is visible in Figure~\ref{fig:label19}.  Here we simulate the Moon directly, and both the Mie scattering halo is seen as well as the attenuation of direct light due to the spatial pattern of the clouds.

\section{Correlated Patterns}

\begin{figure}[htb]
\begin{center}
\includegraphics[width=0.99\columnwidth]{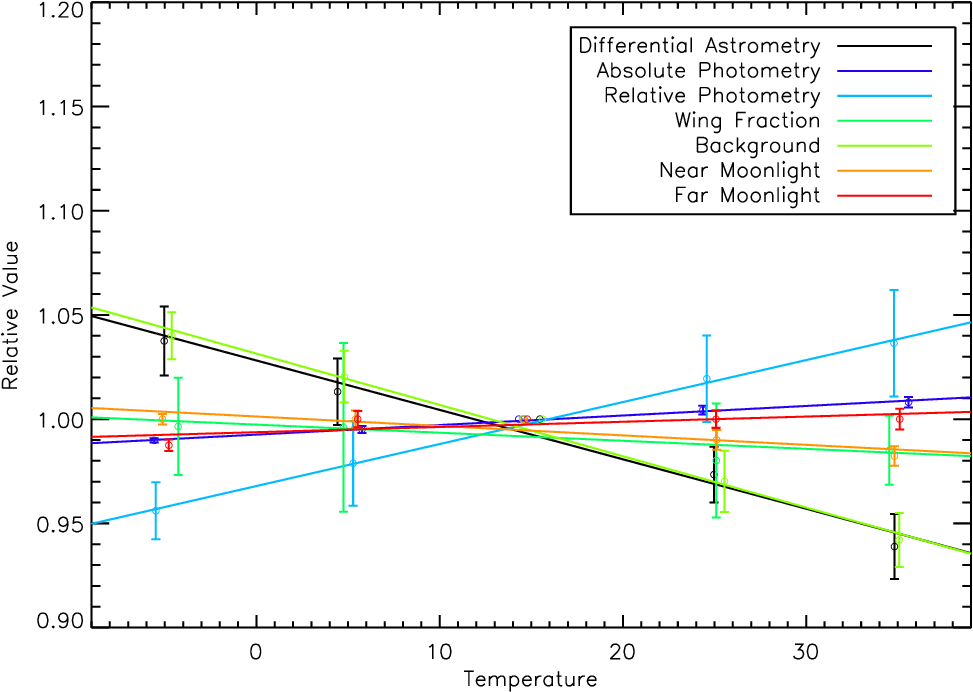}
\end{center}
\caption{\label{fig:label20} Changes in differential astrometry, absolute photometry, relative photometry, PSF wing fraction, background level, near Moonlight background level, and far Moonlight background level as a function of the ground temperature in Celcius.}
\end{figure}

\begin{figure}[htb]
\begin{center}
\includegraphics[width=0.99\columnwidth]{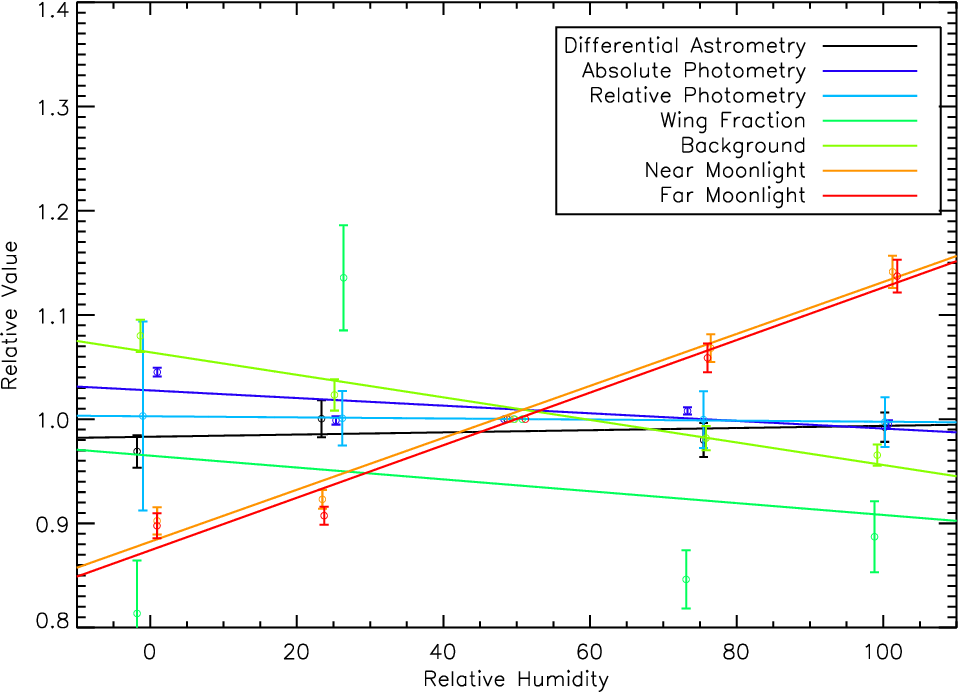}
\end{center}
\caption{\label{fig:label21} Changes in differential astrometry, absolute photometry, relative photometry, PSF wing fraction, background level, near Moonlight background level, and far Moonlight background level  as a function of the relative humidity.}
\end{figure}

\begin{figure}[htb]
\begin{center}
\includegraphics[width=0.99\columnwidth]{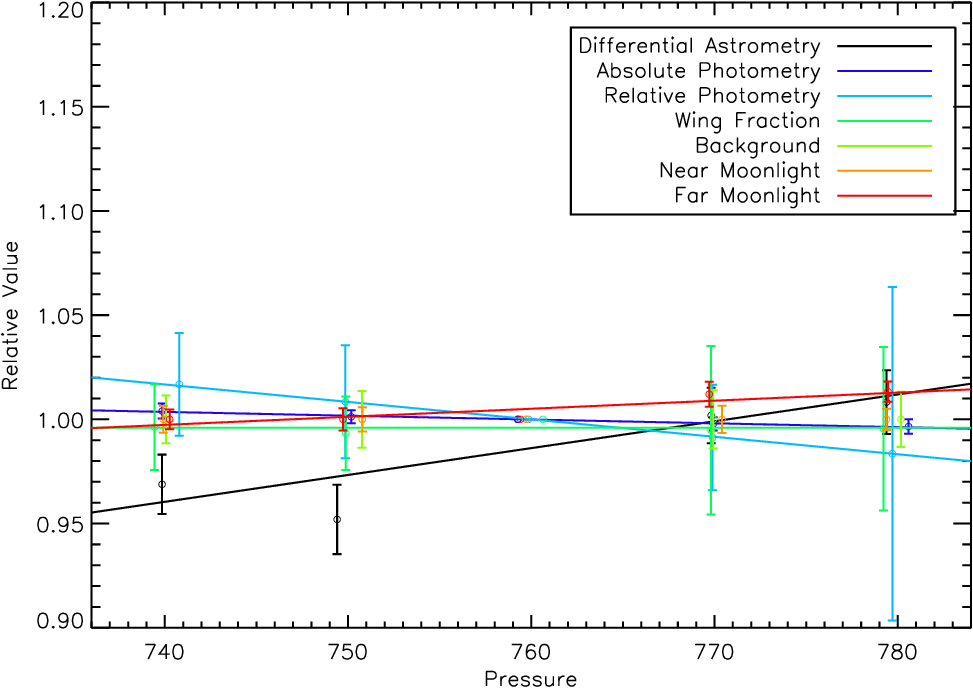}
\end{center}
\caption{\label{fig:label22} Changes in differential astrometry, absolute photometry, relative photometry, PSF wing fraction, background level, near Moonlight background level, and far Moonlight background level  as a function of the ground pressure in mmHg.}
\end{figure}

\begin{figure}[htb]
\begin{center}
\includegraphics[width=0.99\columnwidth]{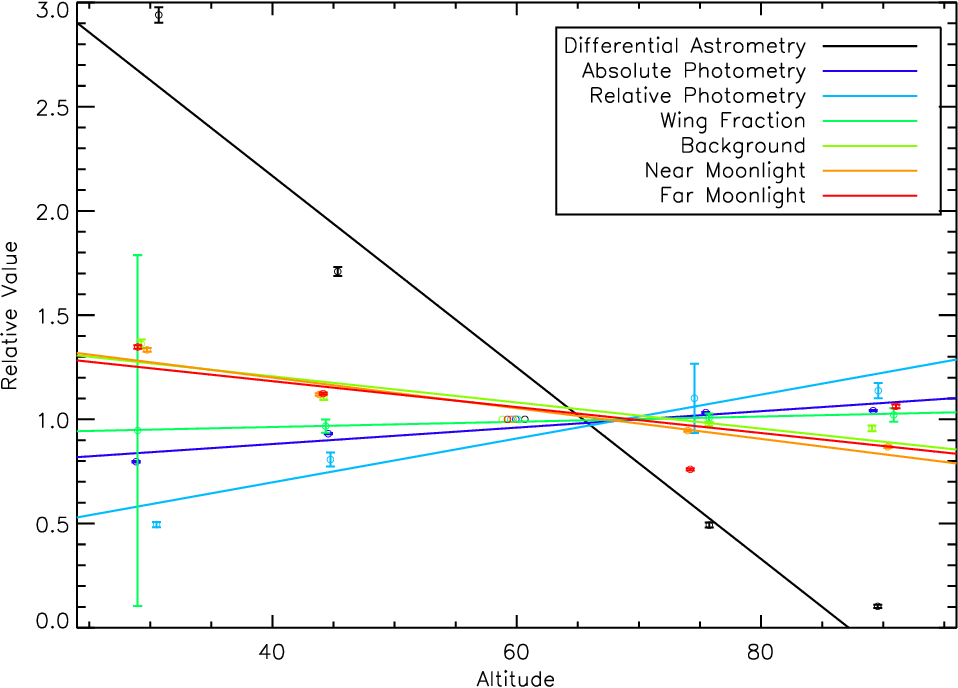}
\end{center}
\caption{\label{fig:label23} Changes in differential astrometry, absolute photometry, relative photometry, PSF wing fraction, background level, near Moonlight background level, and far Moonlight background level  as a function of observation altitude.}
\end{figure}

\begin{figure}[htb]
\begin{center}
\includegraphics[width=0.99\columnwidth]{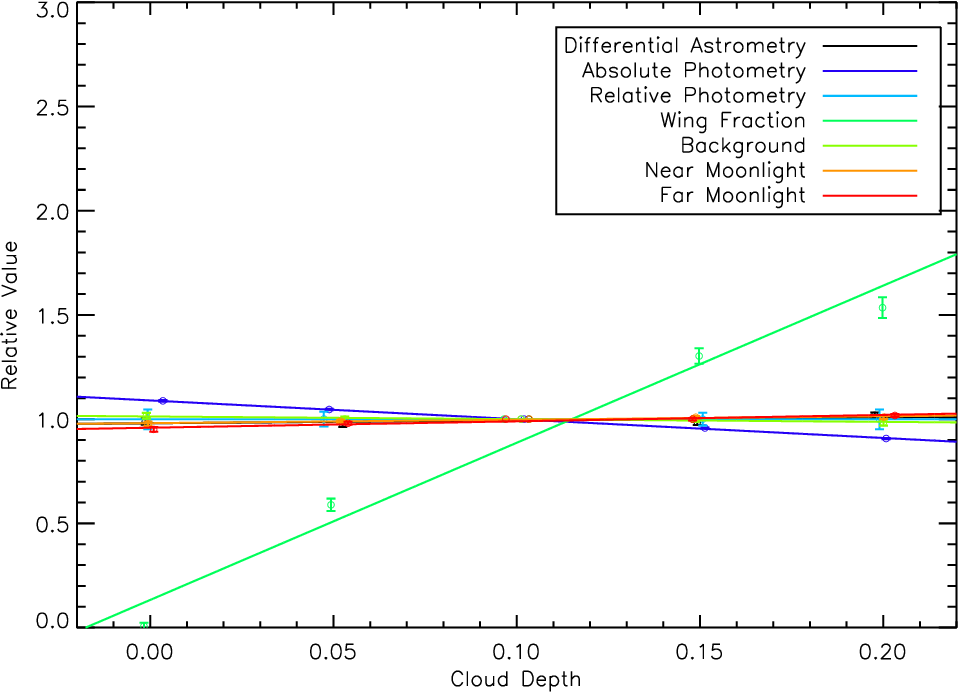}
\end{center}
\caption{\label{fig:label24} Changes in differential astrometry, absolute photometry, relative photometry, PSF wing fraction, background level, near Moonlight background level, and far Moonlight background level as a function of cloud depth in magnitudes.}
\end{figure}

The net effect of the comprehensive implementation is that certain external variables (temperature, pressure, etc.) produce correlated behaviours in the emergent observable properties.  Some of these correlations are well known, but others are more subtle.  To demonstrate this, we perform 150 sets of simulations with the 5 most important external variables modified to 5 different values and all other variables randomly generated.  The five external variables are:  pressure, temperature, humidity, altitude (zenith angle), and cloud depth.  For each of these variables, we varied the quantities between extreme values that would likely be encountered at some point in an observing run.  Thus, temperature was varied between -10 and 40 degrees Celcius, humidity was varied between 0 to 100\%,  pressure was varied between 740 and 780 mmHg, altitude was varied between 30 and 90 degrees, and cloud depth was varied between 0 and 0.2.  In many real world cases, the variation may not be as large (e.g. more cloudy nights might be avoided or switching to a new target at different zenith angles).  However, as we show below the effect on observables is often close to linear, and the variation can then be scaled to a more limited range.  Other secondary variables (e.g. wind patterns, exosphere temperature, aerosol depth, cloud patterns and cover fractions) are marginalized in the 150 sets of simulations.  Therefore, we have $3150=(4 \times 5+1) 150$ total simulations.

For each simulation, we measure several observable properites:  differential astrometry (non-ideal astrometric patterns), absolute photometry (overall transmission), relative photometry (spectrally-dependent transmission), the fraction of light in the PSF wing, overall nominal background level, moon light intensity near the Moon, and Moon light intensity far from the Moon.  We define each metric as follows.  Differential astronometry is measured by using the difference in centroid of two sources with a monochromatic spectral energy distribution at 350 nm and 550 nm.  Absolute photometry is measured by using the ratio of total counts of a source with a 550 nm SED.  Relative photometry is measured by the ratio of total counts from a source with a 350 nm SED and a 550 nm SED.  The fraction of light in the PSF wing is measured by estimating the source flux in rows with a maximum of less than 10 counts with the total flux from a source with a 350 nm SED.  The counts are collected in rows, since the measurement involves highly saturated stars to see the wing pattern and therefore the bleed trail is avoided.  The nominal background is measured by using the median counts using a V band filter.  The Moon background is measured in the same manner.

\begin{deluxetable}{lrrrrrr}
  \tablecolumns{7}
\tablecaption{\label{tab:table2} The systematic percent change of various image observables with the most important environmental variablesfor typical ranges of environmental variables.  The error on each value is about 2\%.  The blank entries in the table correspond with a less than 2\% correlation.}
 \tablehead{
   \colhead{Image Characteristic} &  \colhead{Temperature} & \colhead{Humidity} & \colhead{Pressure} & \colhead{Altitude} & \colhead{Cloud Depth} & \colhead{Residual Variation} \\
   \colhead{(Range) } & \colhead{(0,40$^{\circ}$C)} & \colhead{(0,100\%)} & \colhead{(740,780mmHg)} & \colhead{(30,90$^{\circ}$)} & \colhead{(0,0.2mag)}} 
 \startdata
Differential Astrometry & -10 & 2    & 4   & -284 & -    & 23 \\
Absolute Photometry    & 2    &  -5   & -  & 24   & -18  & 7\\
Relative Photometry     & 8    & -     & -3   & 64   & -       & 59\\
Wing Fraction                 &  -   & -      & -     & -      & 153    & 170 \\
Background                    & -10 & -11  & -     & -40  & -3   & 24  \\
Near Moonlight             &  -   & 24   &  -     & -47  & 3   & 30 \\
Far Moonlight                &  -    & 24   & -    & -28   & 7 & 27 \\
  \enddata
\end{deluxetable}

In Figures~\ref{fig:label20},~\ref{fig:label21},~\ref{fig:label22},~\ref{fig:label23}, and~\ref{fig:label20}, we the change in relative values of the observable quantities with changes in the enviromental variables.  In most cases, the changes are approximately linear.  In Table~\ref{tab:table2} we show the average systematic percent change of observables with the range of the environmental variables with accuracy to 2\%.  These would be the approximate predicted change of observable quantities that an observer would likely encounter with a wide range of conditions.  We produce several well-known correlations that are well-known:   the correlation in altitude with absolute and relative photometry (24\% and 64\%, absorption at higher airmass), the anti-correlation of altitude with differential astrometry (-284\%, atmospheric dispersion), the anti-correlation of background with altitude (-40\%), and the anti-correlation of absolute photometry with cloud depth (-18\%, basic cloud attenuation).  However, several subtle correlations are evident.  One is that astrometry is affected by temperature (-10\%), humidity (2\%), and pressure (4\%) through the wavelength-dependence of the index of refraction of air which is consistent with other studies (\citealt{filippenko1982}, \citealt{stone1996}).  Another is that absolute and relative photometry can be affected by temperature (2\% and 8\%) by modifying the density profile.  This is also seen in the anti-correlation of relative photometry with pressure (-3\%).  Humidity also reduce the absolute photometry through absorption as expected (-5\%).

The PSF wing fraction was quite strongly affected by cloud depth.  This is consistent with the estimate above that most of the PSF wing at arcminute scales is due to water droplets, but the magnitude of the correlation is large (153\%).  It may be possible to then improve photometry by estimating cloud depth by looking at the PSF wing.  The background level was affected by temperature (-10\%), humidity (-11\%), and cloud depth (-3\%) in addition to altitude.  It is difficult to provide a simple explanation for these correlations, since the background level is an emergent property of the calculation.  In addition, scattered moonlight behaves differently than non-lunar background and it varies with the angular distance to the moon.  Here there is a correlation in Lunar background with humidity (24\%, 24\%), anti-correlation with altitude (-47\%, and -28\%), and correlation with cloud depth (3\%, 7\%).  These result implies that in moonless nights background will be better at higher temperatures, higher humidities, and less clouds, but when the Moon is visible lower humidities and more clouds are preferred.

There are also a significant number of subtle correlations of less important variables implied by the physics implementation that are beyond the scope of this work.  The last column in Table~\ref{tab:table2} lists the residual variation of each observable when the five most important environmental variables are held fixed.  This includes other variables implicit in the simulation.
For instance, we found that the twilight light intensity after sunset is most affected by upper atmosphere scattering.  The density in the upper atmosphere is affected by the exosphere temperature, which is related to the solar cycle.  Thus, twilight and sunset intensities should follow the Solar cycle.

\section{Conclusion, Validation, and Future Work}

We presented a comprehensive photon Monte Carlo implementation of the propagation of light through a self-consistent representation of the atmosphere.  The results are fully implemented in v5.7 of PhoSim which is publically available at https://www.phosim.org.  The new capabilities should be sufficient for a number of applications that require realistic images that have complex photometric, astrometric, PSF patterns that depend on observing conditions.  Some results will depend on the input data that includes the climate historical data, the vertical thermodynamic profiles, the aerosol patterns, and the the atomic data which will continue to be validated and studied for non-astronomical applications.  We have shown validation of the cloud model with the structure function and opacity statistics and analytic validation of differential chromatic refraction for low zenith angles.  Numerical validation of PhoSim is extensive and includes photon positional errors to thousandths of pixels and transmission accuracies to one part in a million.  The emergent predicted results of the PSF wing, background levels at various sites, transmission variation, and the correlations with environmental variables can be studied by the astronomical community in a number of different ways.   Continued validation by the entire community will build trust in the capabilities, and improve and make the results more robust.  Future work will refine the input data as future studies come to a greater understanding of atmospheric details, and the code is constructed so users can input their own atmosphere parameters to custom match specific observations or explore dependencies.  The capabilities included in this work should lead to a more thorough understanding of the effect of the atmosphere on image properties and many diverse measurements in astronomy.

\acknowledgments
We thank an anonymous referee for many helpful comments.  JRP acknowledges support from Purdue University.   JRP acknowledges innumerable important and helpful conversations with J. G. Jernigan.  We acknowledge critical conversations about layer-wise absorption with M. Wood-Vasey and cloud photometry statistics with C. Stubbs.  This work greatly benefited from the AERONET, HITRAN, CIESEN, KGRD, and NCEP/NCAR databases.

\end{document}